\documentclass{aa}  
\usepackage{graphicx}
\usepackage{txfonts}
\usepackage{multirow}
\usepackage{xcolor}

\begin{document}

   \title{Numerical simulations for the SAXO+ upgrade: Performance analysis of the adaptive optics system}


   \author{C. Goulas \inst{1}\fnmsep\thanks{\email{charles.goulas@obspm.fr}}
          \and R. Galicher \inst{1}
          \and F. Vidal \inst{1}
          \and J. Mazoyer \inst{1}
          \and F. Ferreira \inst{1}
          \and A. Sevin \inst{1}
          \and A. Boccaletti \inst{1}
          \and E. Gendron \inst{1}
          \and C. Béchet \inst{2}
          \and M. Tallon \inst{2}
          \and M. Langlois \inst{2}
          \and C. Kulcsár \inst{3}
          \and H-F. Raynaud \inst{3}
          \and N. Galland \inst{3}
          \and L. Schreiber \inst{4, 7}
          \and I. Bernardino Dinis \inst{5}
          \and F. Wildi \inst{5}
          \and G. Chauvin \inst{6}
          \and J. Milli \inst{7}
          }

   \institute{LESIA, Observatoire de Paris, Université PSL, Université Paris Cité, Sorbonne Université, CNRS, 5 place Jules Janssen, 92195 Meudon, France
             \and
             CRAL, CNRS, Université Claude Bernard Lyon 1, ENS de Lyon
             \and
             IOGS, CNRS, Laboratoire Charles Fabry, Université Paris-Saclay
             \and
             INAF, Osservatorio di Astrofisica e Scienza dello Spazio di Bologna
             \and
             Dept. of Astronomy, University of Geneva
             \and
             Laboratoire J.-L. Lagrange, CNRS, OCA, Université Côte d’Azur
             \and
             IPAG, CNRS, Université Grenoble Alpes
             }

   \date{Received September 15, 1996; accepted March 16, 1997}

 
  \abstract
   {SPHERE, operating at the VLT since 2014, is currently one of the high-contrast instruments with a higher performance. Its adaptive optics system, known as SAXO, will be upgraded to SAXO+, which features the addition of a second stage of adaptive optics. This stage will use a near-infrared pyramid wavefront sensor to record images of fainter exoplanets around redder stars.}
   {In this work, we compare the performance of SAXO and SAXO+. We look for the optimal values of the key system parameters of SAXO+ for various science cases and turbulence conditions.}
   {We performed numerical simulations using COMPASS, an end-to-end adaptive optics simulation tool. We simulated perfect coronagraph images of an on-axis point source, and we minimized the residual starlight intensity between~3 and $5\,\lambda / D$ as a performance criterion. The explored parameter space includes science cases (described by magnitude in~G and~J bands), turbulence conditions (seeing and coherence time), and key system parameters (first and second stage gains, first and second stage frequencies, pyramid modulation radius, pyramid modal gains optimization).}
   {In every science case and turbulence condition, SAXO+ reduces the residual starlight intensity inside the correction zone of the second stage by a factor of ten compared to SAXO. The optimal first stage gain is lower for SAXO+ than for SAXO alone. We quantified the gain in performance of SAXO+ when changing the second stage frequency from 2\,kHz to 3\,kHz, and we conclude that 2\,kHz may be sufficient for most realistic conditions. We give the optimal first stage gain as well as the first and second stage frequencies for every seeing, coherence time, and science case. Finally, we find that a $2\,\lambda_{\mathrm{WFS}} / D$ pyramid modulation radius is a good trade-off between performance and robustness against varying turbulence conditions.}
   {This study shows that the future SAXO+ system will outperform the current SAXO system in all studied cases.}


   \maketitle

\section{Introduction}

The aim of high-contrast imaging is to detect light emitted or reflected by the near surroundings of stars. This allows for spectroscopic and polarimetric characterization of circumstellar disks and exoplanet atmospheres. However, among the 5000 exoplanets discovered so far, less than 1\% have been directly imaged.\footnote{\url{https://exoplanet.eu}} Obtaining an image of an exoplanet is challenging for two reasons. First, the exoplanet is much fainter than its host star. The planet-to-star luminosity ratio ranges from $10^{-4}$ for the brightest young Jupiters to $10^{-10}$ for an exo-Earth in visible and near-infrared. Secondly, an exoplanet can be very close to its host star, at angular separations lower than 1". Coronagraphs can block the starlight, letting the exoplanet light reach the detector. However, coronagraphs require an aberration-free wavefront. Hence, for a ground-based instrument, an adaptive optics (AO) system is needed to correct for Earth's atmospheric turbulence. State-of-the-art exoplanet imagers include GPI at Gemini South \citep{Macintosh_2018}, Clio2/MagAO at the \textit{Magellan} telescope \citep{Sivanandam_2006, Close_2010}, SCExAO at Subaru \citep{Jovanovic_2015}, and SPHERE at VLT \citep{Beuzit_2019}. GPI and MagAO are also being upgraded to GPI 2.0 \citep{Chilcote_2022} and MagAO-X \citep{Males_2022}, respectively. These instruments currently allow the detection of massive and young Jupiters, which mainly emit in the near-infrared.

SPHERE has been observing at the VLT since 2014. Its AO system, called SAXO \citep{Fusco_2014}, provides a corrected beam to three coronagraph instruments, IRDIS \citep{Dohlen_2008}, IFS \citep{Mesa_2015}, and ZIMPOL \citep{Schmid_2018}. SAXO contains a $40\times40$ Shack-Hartmann \citep[SH;][]{Shack_1971} wavefront sensor (WFS) working in visible light (500-900 nm), a $41\times41$ high-order deformable mirror (HODM), and a fast tip-tilt mirror~\citep{baudoz2010}. The starlight intensity (normalized by the maximum of the non-coronagraph image) in the raw images of SPHERE reaches $10^{-4}$ at 300 mas \citep{Cantalloube_2019} before post-processing techniques. 

In exoplanetary science, some of the most exciting recent discoveries have been detected around very young stars showing the formation process of planets~\citep[e.g., the multiple system orbiting PDS 70, ][]{keppler2018_DiscoveryPlanetarymassCompanion, haffert_2019}. However, these young stars are very red and at the limit of the current AO capabilities. Due to the WFS sensitivity, the performance of SAXO drops for stars fainter than magnitude 12 in the V band \citep{Milli_2017, Jones_2022}. Furthermore, even for bright stars, the temporal error of the AO loop becomes significant when the coherence time of the atmosphere is below 3~ms. When this occurs, a halo of stellar light, known as the wind-driven halo, may appear in the coronagraph images, increasing the starlight in the final image \citep{Cantalloube_2020}. To tackle these limitations and improve detection capabilities, the SPHERE consortium proposes an upgrade of SAXO called SAXO+. This upgrade must answer three key scientific requirements defined by \citet{Boccaletti_2020}: reach the young giant planet population down to the snow line, observe fainter and lower mass stars, and enhance the characterization of exoplanetary atmospheres. SAXO+ will create deeper contrasts at smaller angular separations for all stars and allow for the observation of fainter and redder stars than what SPHERE currently permits.

\begin{figure}
    \centering
    \includegraphics[width=0.9\hsize]{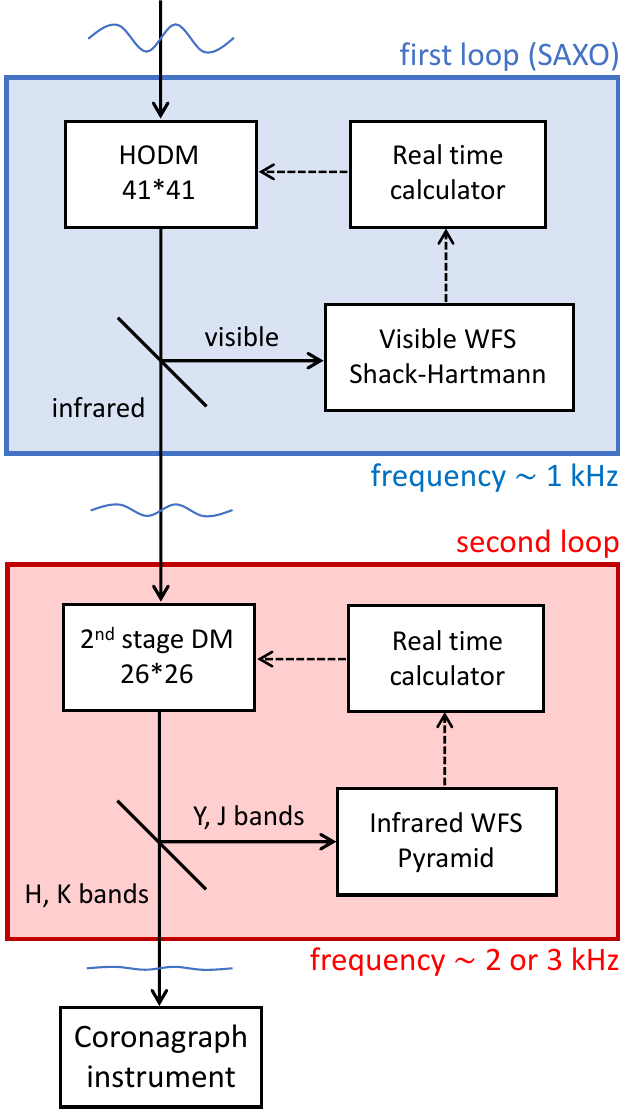}
    \caption{Current design of SAXO+. The first stage maximum frequency is 1.38 kHz, and the second stage maximum frequency is 3 kHz. There is also a mode for bright targets where only the Y band is used by the pyramid WFS, but we do not simulate this case.}
    \label{fig:design}
\end{figure}

To achieve these goals, the SAXO+ design includes a second stage AO downstream of the current SAXO stage, as described in Fig. \ref{fig:design}. SAXO+ is part of the road map of the European Southern Observatory (ESO) instrument PCS/ELT \citep{Kasper_2021} as a technical demonstrator of a two-stage AO for high-contrast imaging. In SAXO+, the second stage is faster, with a maximum speed of 3 kHz, in order to address the temporal error of SAXO. The wavefront sensing is done with near-infrared light, at 1.2 \textmu m, with a pyramid WFS \citep[PWFS,][]{Ragazzoni_1996} that is more sensitive than the SH WFS.
The SAXO+ consortium chose the $28\times28$ kilo-DM from Boston Micromachines. As the actuators on the edge of the DM are not fully controllable, there are actually 26 actuators across the telescope pupil diameter. Several solutions are possible for the RTC as well as for how to control the different elements in a two-stage AO. In the integrated case, one RTC receives inputs from the WFSs and controls the two DMs. This solution will not be used in SAXO+. In the stand-alone case, the two loops are independent, meaning there is one RTC for each loop. This is the baseline solution chosen for SAXO+. In a third case that has been investigated by the SAXO+ consortium, the control algorithms of the second stage account for the closed-loop telemetry of the first stage. In this paper, we only consider the stand-alone solution with two independent loops. The cascade AO-system design has shown encouraging results in simulations by \citet{Cerpa-Urra_2022} and experimentally on an optical test bed by \citet{NDiaye_2023}.

In this paper, we study the SAXO+ system in numerical simulations (Sect. \ref{sec:framework}). The results aim to contribute to the trade-offs and optimization in the instrument design. In Sect. \ref{sec:param_study}, we show the improvement expected from SAXO+ compared to SAXO in terms of residual starlight intensity in the coronagraph image. We study the impact of the gain of the first stage on the SAXO+ performance and on the displacement of second stage DM actuators. We quantify performance as a function of the second stage frequency. Then, we give the optimal system parameters (first stage gain, first and second stage frequencies) for all simulated turbulence conditions and science cases. Finally, we present a specific study about the PWFS modulation radius.

\section{Simulation framework}
\label{sec:framework}

To design a simulation of the SAXO+ system, one needs to simulate the atmospheric turbulence that causes aberrations, the SAXO+ AO system that corrects some of the aberrations, and the coronagraph image in which the performance is measured. In Sect.~\ref{subsec:simtool}, we present the simulation tool. In Sect.~\ref{subsec:science_case}, we describe the science cases identified by the SAXO+ consortium. In Sect.~\ref{subsec:tubu}, we describe the numerical simulations of the atmosphere turbulence. In Sect.~\ref{subsec:aosyst}, we give the assumptions we make about several hardware and software subsystems of SAXO+. In Sect.~\ref{subsec:coro}, we explain how we simulated the coronagraph images, and we define the normalized intensity that we used as a metric to probe the performance in this paper.

\subsection{Simulation tool}
\label{subsec:simtool}
The numerical simulations were performed with COMPASS, an end-to-end AO simulation tool \citep{Gratadour_2016}. COMPASS simulates atmospheric turbulence, the telescope pupil, AO subsystems (SH and pyramid WFS, DM, control algorithm), and image formation. The models of each of these components are described on the official website of the COMPASS tool.\footnote{\url{https://compass.pages.obspm.fr/website/}} Light propagation from a pupil plane to a focal plane, and vice versa, is based on Fourier optics.

As described in \citet{Goulas_2023}, we upgraded the COMPASS platform with key features for the SAXO+ simulations. Thus, it can now include a SH WFS and a pyramid WFS in cascaded simulations, handle the temporal asynchrony of a two-stage AO system, and compute coronagraph images by simulating electric field propagation through either a coronagraph composed of pupil and focal plane masks or a perfect coronagraph \citep{Cavarroc_2006}.

In this paper, the parameters of the simulations (hardware, software, turbulence) were set using knowledge on the current system (SAXO, VLT atmosphere conditions) and on the scientific requirements \citep{Boccaletti_2020}. All parameters are summarized in Table \ref{param_1}. We studied the impact of the key parameters (the ones with several values in the table) on the coronagraph performance.

\begin{table}
    \renewcommand{\arraystretch}{1.12}
    \centering
    \caption{Simulation parameters.}
    \begin{tabular}{ll}
    \hline
        \multicolumn{2}{c}{\textbf{Turbulence}} \\
    \hline
        profile & ESO 35 layer median \\
        outer scale & $L_0$ = 25 m \\
        seeing* & $s$ = 0.5, 0.8, 1.2 arcsec \\
        coherence time* & $\tau_0$ = 1, 3, 5, 7 ms \\
    \hline
        \multicolumn{2}{c}{\textbf{Telescope}} \\
    \hline
        diameter & $D = 8$ m \\
        entrance pupil & VLT pupil with spiders \\
    \hline
        \multicolumn{2}{c}{\textbf{First stage}} \\
    \hline
        \multicolumn{2}{c}{SH WFS} \\
        wavelength & $\lambda_{\mathrm{WFS}}$ = 700 nm \\
        subapertures & 40 $\times$ 40 \\
        readout noise & 0.1 electrons/pixel \\
        \multicolumn{2}{c}{Deformable mirror} \\
        geometry & 41 $\times$ 41 + tip-tilt mirror \\
        modal basis & 800 KL modes \\
        \multicolumn{2}{c}{Control loop} \\
        reconstruction matrix & least squares method \\
        command law & integrator + scalar gain \\
        total delay & 2.15 ms \\
        \multirow{2}{*}{loop gain*} & $g_1$ = 0, 0.01, 0.05, \\
         & 0.1, 0.2, 0.3, 0.5 \\
    \hline
        \multicolumn{2}{c}{\textbf{Second stage}} \\
    \hline
        \multicolumn{2}{c}{Pyramid WFS} \\
        wavelength & $\lambda_{\mathrm{WFS}}$ = 1.2 \textmu m \\
        subapertures & 50 $\times$ 50 \\
        readout noise & 0.8 electrons/pixel \\
        modulation radius* & 0, 1, 3, 5, 10 $\lambda_{\mathrm{WFS}} / D$ \\
        \multicolumn{2}{c}{Deformable mirror} \\
        geometry & 26 $\times$ 26 \\
        modal basis & 400 KL modes \\
        \multicolumn{2}{c}{Control loop} \\
        reconstruction matrix & least squares method \\
        command law & modal integrator \\
        optical gain compensation* & CLOSE algorithm, none \\
        total delay & 2 sampling frames \\
    \hline
        \multicolumn{2}{c}{\textbf{Image formation}} \\
    \hline
        type of coronagraph & perfect \\
        wavelength & $\lambda$ = 1.67 \textmu m \\
        exposure time & 3 s \\
    \end{tabular}
    \tablefoot{The asterisk indicates the variable parameters of this study. The entries without an asterisk are fixed parameters.}
    \label{param_1}
\end{table}

\subsection{Science cases}
\label{subsec:science_case}

\begin{table*}
    \renewcommand{\arraystretch}{1.15}
    \centering
    \caption{Science cases and simulated frequencies.}
    \begin{tabular}{|ccccc|c|c|ccc}
    \hline
        \multirow{2}{*}{Science case} & \multirow{2}{*}{G mag} & \multirow{2}{*}{J mag} & SH flux & pyramid flux & \multicolumn{2}{c|}{simulated frequencies} \\
        \cline{6-7}
         & & & [ph-e/subap/ms] & [ph-e/pixel/ms] & 1st stage [Hz] & 2nd stage [Hz] \\
    \hline
        bright-1 & 5.5 & 5.2 & 411 & 146 & \multirow{2}{*}{500, 1000} & \multirow{6}{*}{1000, 2000, 3000} \\
        bright-2 & 7.6 & 7.2 & 59.3 & 23.1 & & \\
        \cline{6-6}
        bright-3 & 9.6 & 7.8 & 9.41 & 13.3 & 250, 500, 1000 & \\
        \cline{6-6}
        red-1 & 11.9 & 8.5 & 1.13 & 6.98 & \multirow{2}{*}{50, 250, 500} & \\
        red-2 & 12.8 & 10.1 & 0.492 & 1.60 & & \\
        \cline{6-6}
        red-3 & 14.5 & 10.1 & 0.130 & 1.60 & \multirow{2}{*}{50, 250} & \\
        \cline{7-7}
        red-4 & 15.3 & 11.4 & 0.0492 & 0.483 & & 500, 1000, 2000 \\
        \cline{6-7}
        red-5 & 16.8 & 12.5 & 0.0124 & 0.175 & 10, 50, 250 & 100, 250, 500 \\
        \hline
    \end{tabular}
    \label{param_2}
\end{table*}

The SAXO+ consortium defined eight science cases, given in Table \ref{param_2} \citep{Schreiber_2023}. For each case, the science target is also the AO guide star, and we provide the G and J magnitudes, as it better captures the spectral range we are dealing with in SAXO and SAXO+. We deduced the photon flux on the SH WFS from the~G magnitude $m_G$ and the photon flux on the PWFS from the~J magnitude $m_J$. The brightest target is "bright-1," with $m_G=5.5$ and $m_J=5.2$. The magnitudes at the two bands increase until the faintest case, "red-5." As suggested by their name, the five faintest targets are also red stars, with $m_G-m_J=2.4$ for the "red-1" case and up to $m_G-m_J=4.3$ for red-5. For the red cases, the photon flux on the SH WFS is lower than the photon flux on the PWFS. The last two columns show the frequencies of each stage that we tested in this paper. For the brightest stars, the photon flux level is high enough to run the system at the maximum speed. As the number of photons on the WFS decreased, we slowed down the first or second stage to ensure there was at least one photo-electron per frame and per subaperture for the SH or per pixel for the PWFS.

\subsection{Turbulence}
\label{subsec:tubu}
The simulation of the atmosphere is based on the ESO 35 layer "median" profile for which we call~$v_{\mathrm{median}}$ the effective wind speed \citep{Vidal_2019}. Three seeing conditions were simulated: a good seeing of~0.5", an average seeing of~0.8", and a poor seeing of~1.2". We simulated four values of coherence time: 1, 3, 5, and 7 ms. We recall that the median coherence time at Paranal is 4.5 ms and that \citet{Cantalloube_2020} showed that the wind-driven halo dominates in the raw images of SPHERE when the coherence time of the atmosphere is below 3 ms. To simulate a specific coherence time $ \tau_0 $, we computed the effective windspeed $ v = 0.314\,r_0 / \tau_0 $, where $r_0$ is the Fried parameter. We then multiplied the windspeed of each layer of the ESO 35 layer profile by $ v / v_{\mathrm{median}}$.

\subsection{Adaptive optics system}
\label{subsec:aosyst}
\subsubsection{Wavefront sensor}

In the current SAXO system, the SH WFS is equipped with a square spatial filter in a focal plane to improve aliasing rejection. The field of view can be 0.82", 0.89", 1.07" wide or the full field view, which is selected by the telescope operator depending on the seeing \citep{Sauvage_2014}. In the simulations, we chose the 0.82" diameter for 0.5" seeing, the 0.89" diameter for 0.8" seeing, and the open position for 1.2" seeing, which are the typical operating choices. To measure the position of the spot on the SH WFS, the current system implements the thresholded and weighted center of gravity \citep{Sauvage_2014}. We used the same algorithm in the simulations.

The pyramid WFS is well known for its dependency on the properties of the incoming wavefront \citep[in particular sensitivity and linearity, see][]{Verinaud_2004}. We decided to use the recently developed "full pixel method" \citep{Clergeon_2014, Deo_2018} for the pyramid measurement. This method is more robust against pyramid misalignments and manufacturing defects than the traditional gradient sensing scheme. The pyramid measurement vector contains the signal of each valid pixel, normalized by the mean of the valid pixels. The valid pixels on the pyramid detector include every pixel in which at least 0.1\,\% of the surface is inside the pupil. The modulation radius of the PWFS was set to 3 $\lambda_{\mathrm{WFS}} / D$, except in paragraph \ref{subsec:modu}.

The PWFS wavelength is 1.2 \textmu m, according to the system design choice. As the science cases of SAXO+ are mostly in the H band and the SH WFS uses visible photons, the Y and J bands are dedicated to the PWFS. A recent study has suggested that unlike the SH, it is a wise choice to slightly oversample the geometry of the pyramid subapertures compared to the pattern of the DM actuators across the pupil \citep{Vidal_2018}. Furthermore, in SAXO+ the second stage DM has less actuators than the first stage DM, and -- although this is not part of this study -- it is still considered by the project to use the PWFS to control the first stage DM at a later point. As such, we decided to simulate a 50 $\times$ 50 sampling to eventually accommodate for the control of the first stage DM if required (41 $\times$ 41 actuators).

The reference slopes of the SH WFS and the reference measurement vector of the PWFS are defined by a flat wavefront (no aberration). Non-common path aberrations (NCPAs) are not taken into account.

\subsubsection{Deformable mirror}
\label{subsubsec:dm}

For the first stage, we simulated the SAXO deformable mirror of $41\times41$ actuators and the tip-tilt mirror. The dead actuators of the current HODM installed on SPHERE were not simulated in this paper, which aims at choosing the main parameters of the SAXO+ system. Once these parameters are chosen, we will add the known dead actuators of the current HODM to study their impact on the final performance, but this is out of the scope of this paper.

The second stage DM is composed of $28\times28$ actuators. The choice of the number of actuators resulted from a trade-off between technical requirements (update rate, stroke), cost, and market supply. The pupil was sized to $26\times26$ actuators because the optical aperture recommended by Boston Micromachines does not include the actuators at the edge, which are not fully controllable. The optical stroke was~11\,\textmu m.

For each of these deformable mirrors, we computed simplified Karhunen-Loève (KL) modes, as described by \citet{Bertrou_2022}. We call $\vec{B}_1$ the modal basis matrix of the first stage DM and $\vec{B}_2$ the modal basis matrix of the second stage DM. The columns of $\vec{B}_1$ and $\vec{B}_2$ contain the modes of the basis expressed in the actuator space. The highest order modes are filtered by truncation. We used 800 modes on the first stage and 400 modes on the second stage.

\subsubsection{Command matrices}

The modal interaction matrices $\vec{D}_1$ and $\vec{D}_2$ were calibrated without photon or detector noise under infinite flux conditions. We note that $\vec{D}_1$ contains, in column, the slopes vector of the SH WFS measured by pushing each mode of $\vec{B}_1$ one after the other with~1\,nm RMS amplitude. Similarly, $\vec{D}_2$ contains, in column, the measurement vector of the PWFS obtained by pushing on each mode of $\vec{B}_2$ (one after the other with a~1\,nm RMS amplitude). The same pyramid modulation was applied during the interaction matrix calibration procedure and while closing the loop. 

The modal command matrices $\vec{R}_1$ and $\vec{R}_2$ are the generalized inverse of the modal interaction matrices $\vec{D}_1$ and $\vec{D}_2$, respectively. Mathematically speaking, with $j = 1$ or $2$ respectively for the first and second stage, \begin{equation}
    \vec{R}_j = (\vec{D}_j^\intercal \vec{D}_j)^{-1}\vec{D}_j^\intercal.
\end{equation}

\subsubsection{Command law}

For the first stage, the command law is a temporal integrator with a scalar gain, $g_1$, based on the following recurrence equation:
\begin{equation}
    \vec{c}_1[i+1] = \vec{c}_1[i] - g_1 \vec{B}_1 \vec{R}_1 \vec{m}_1[i]
    \label{eq:integrator1},
\end{equation}
where $\vec{c}_1$ is the command vector containing the DM voltages of the first stage, $\vec{m}_1$ is the SH slopes vector, and $i$ is the iteration of the recurrence. With the gain $g_1$, we could handle the temporal optimization of the loop.

For the second stage, we additionally used a vector of modal gains $\vec{g}_{\mathrm{m}}$:
\begin{equation}
    \vec{c}_2[i+1] = \vec{c}_2[i] - g_2 \vec{B}_2 \vec{g}_{\mathrm{m}} \vec{R}_2 \vec{m}_2[i]
    \label{eq:integrator2},
\end{equation}
where $\vec{c}_2$ is the command vector containing the DM voltages of the second stage and $\vec{m}_2$ is the pyramid measurement vector.

As the amplitude of a mode increases, the nonlinearity of the PWFS results in a sensitivity loss. This phenomenon can be described in each mode by an optical gain between zero and one, which we call $\alpha_k$ for the mode number $k$. Thus, the effective gain of the loop for this mode is $g_2 \alpha_k g_k$, with $g_k$ as the $k$ component of the modal gain vector $\vec{g}_{\mathrm{m}}$. The $\alpha_k$ optical gains describe a physical optical effect of the pyramid, and they are included in the PWFS measurement $\vec{m}_2$. That is why they do not explicitly appear in Eq.~\ref{eq:integrator2}. Several methods were developed to calibrate the optical gains and to compensate for them during observations \citep{Korkiakoski_2008, Deo_2018, Esposito_2020, Chambouleyron_2021, Agapito_2023}.

In this paper, we use the CLOSE algorithm \citep{Deo_2021}, a real-time optimization of the modal gains $g_k$. CLOSE retrieves information about the transfer function of Eq. \ref{eq:integrator2} from the closed loop pyramid measurements $\vec{m}_2$. Then, with a model of the transfer function, CLOSE estimates the optimal gain of the loop and adjusts the value of $g_k$ so that the effective gain $g_2 \alpha_k g_k$ is equal to the estimated optimal gain.

The total delay of the first stage corresponds to the value measured on the real SAXO system (1.56 ms, \citealt{Cantalloube_2020}). For the second stage, we made an estimation of the time required for the PWFS camera readout, RTC computational time, electronics communication, and DM surface update, which is about 300 \textmu s. Thus, we set in this framework a total delay of two sampling frames.

The command laws in equations \ref{eq:integrator1} and \ref{eq:integrator2} are independent in this paper. In particular, the measurement of the pyramid is not used to control the first stage DM. Although an integrator with a scalar gain is a robust control scheme, this approach is quite conservative, with the exception of the CLOSE algorithm. As a two-stage AO system, SAXO+ might benefit from more ingenious control techniques. Some techniques are currently being studied for implementation in the second stage RTC: linear quadratic Gaussian regulator \citep{Sivo_2014}, disentangled cascaded AO \citep{Galland_2023}, inverse problem approach \citep{Bechet_2023}, and data-driven control \citep{Dinis_2022}. However, those studies are beyond the scope of this work. Our numerical simulations used a baseline controller, and the results of our parametric study will serve as a reference for future comparisons with enhanced control laws.

\subsection{Simulated coronagraph}
\label{subsec:coro}
Since SAXO and SAXO+ are built to provide a corrected beam to the coronagraph instruments (IRDIS/IFS), we simulated long-exposure coronagraph images. These instruments aim at optically attenuating the starlight so that faint sources can be detected (exoplanets, circumstellar disks, or any source in the vicinity of a bright source). For all coronagraph images in this paper, we considered no photon noise, no detector noise, and no NPCAs between the coronagraph channel and the AO WFS channel. Hence, we only studied the impact of the AO performance on the coronagraph image as a function of the AO system parameters (Table~\ref{param_1}). And the only criteria we used to rate the AO performance was the residual starlight intensity in the final coronagraph image. The lower the residual starlight intensity is, the better the performance of the AO.

We could have chosen to simulate the current coronagraphs that are installed in SPHERE or the ones that are foreseen for SAXO+. In both cases, the coronagraph image could have been limited by diffraction patterns that cannot be corrected using the pyramid or SH WFS (spiders, central obstruction, coronagraph mask effects). For example, \citet{Potier_2022} demonstrated that the diffraction patterns (as well as NPCA speckles) can be minimized using a focal plane wavefront controller. After such a correction, the limitation in the coronagraph image comes from the AO residual halo. Therefore, we can write that the AO has to minimize the turbulence halo and that the focal plane wavefront controller has to minimize the diffraction and speckle pattern.

In this paper, we did not include any technique of focal plane correction and concentrated on studying the AO performance. Hence, we wanted to measure the residual AO halo behind a coronagraph with no bias from the coronagraph (i.e., the diffraction pattern). That is why we assumed a perfect coronagraph \citep{Cavarroc_2006}. We used a point spread function (PSF) to refer to the non-coronagraph image, that is to say, the image recorded without the coronagraph but with AO-corrected aberrations. Then, to simulate the perfect coronagraph, we computed the pupil plane electric field associated with the post-AO phase and removed the mean value of this field. Finally, the perfect coronagraph image is the square modulus of the Fourier transform of this field. The non-coronagraph image, which is the PSF, was computed in the same way but without the mean subtraction.
The coupling of the AO and the focal plane wavefront control will be studied in a future paper.

\subsection{Criteria of performance}
\label{subsec:crit}

All coronagraph images were normalized by the maximum intensity of the PSF. The normalized intensity $\eta(\vec{x})$ at a given position $\vec{x}$ in the coronagraph focal plane can be expressed as
\begin{equation}
    \eta(\vec{x}) = \frac{I(\vec{x})}{\max(\mathrm{PSF})}
    \label{eq:metric},
\end{equation}
where $I(\vec{x})$ is the intensity of the coronagraph image before normalization. The metric we used is the azimuthal average $\mu$ of the normalized intensity $\eta$ as a function of angular separation from the optical axis. At the angular separation $s ~\lambda / D$ ($s$ a real number), $\mu$ can be written as
\begin{equation}
    \mu (s) = \frac{1}{\mathcal{A}_s}\int _{\mathcal{A}_s} \eta(\vec{x}) \, \mathrm{d}\vec{x},
\end{equation}
where $\mathcal{A}_s$ is the ring-shaped area between the angular separation $(s - 0.5) ~\lambda / D$ and $(s + 0.5)~\lambda / D$ from the star. We also computed the standard deviation of the normalized intensity over the same area.

There are several options regarding how to use the normalized intensity for a performance criteria. For instance, we can compare the normalized intensity curves of $\mu$ with respect to the angular separation. However, for massive comparison of parameters, it is more suitable to reduce the performance criteria to one number. We defined the criterion $C(s_1, \,s_2)$ as the average of the normalized intensity $\eta$ between two angular separations $s_1~\lambda / D$ and $s_2~\lambda / D$\,:
\begin{equation}
    C(s_1, \,s_2) = \frac{1}{\mathcal{A}_{s_1,\,s_2}}\int _{\mathcal{A}_{s_1,\,s_2}} \eta(\vec{x}) \, \mathrm{d}\vec{x},
\end{equation}
where $\mathcal{A}_{s_1,\,s_2}$ is the ring-shaped area between the angular separation $(s_1 - 0.5)~\lambda / D$ and $(s_2 + 0.5)~\lambda / D$ . In this paper, as a baseline, we used $s_1 = 3$ and $s_2 = 5$, as one of the main science requirements of SAXO+ is to detect exoplanets as close as possible to a star \citep{Boccaletti_2020}. In several figures, the Strehl ratio~(SR) is given for information~\citep{Mahajan_1983}.

\section{Results: Parametric study}
\label{sec:param_study}
This section presents a parametric study of the AO performance of SAXO+ and SAXO. We first show coronagraph images for three typical cases in Sect.~\ref{subsec:images}. In Sect.~\ref{subsec:1st_gain}, we show that the gain of the first stage (SAXO) has to be adapted when the second stage is used in order to ensure an optimal performance. In Sect.~\ref{subsec:2nd_freq} and~\ref{subsec:opt_param}, we explore the following parameters: seeing, coherence time, first stage gain, first and second stage frequencies, and target brightnesses. In total, we simulated 3276 combinations of these parameters (see Table~\ref{param_1}). In Sect.~\ref{subsec:opt_param}, we give the optimal AO parameters for each observing condition (target and weather conditions). In Sect.~\ref{subsec:modu}, we focus on the modulation radius of the pyramid and the impact of the modal gain optimization performed by the CLOSE algorithm.

\subsection{Coronagraph images}
\label{subsec:images}

\begin{figure}
    \centering
    \includegraphics[width=\hsize]{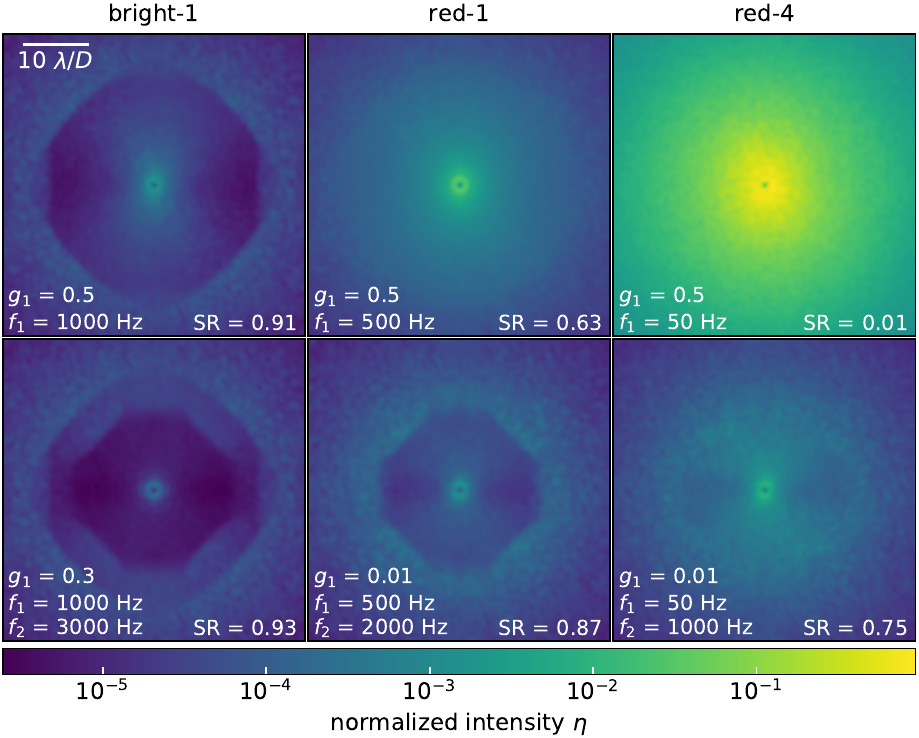}
    \caption{Coronagraph images after SAXO (first row) and SAXO+ (second row) in the science cases bright-1 (left), red-1 (center), and red-4 (right). Conditions: $s = 0.8$", $\tau_0 = 3$ ms. Imaging wavelength: $\lambda = 1.67$ \textmu m. The system parameters (gain and frequency of the first stage, respectively $g_1$ and $f_1$, and frequency of the second stage $f_2$) were optimized to minimize the angular distance residuals between 3 and 5 $\lambda / D$ and are given in the lower-left corner of each image. The PWFS modulation radius is 3 $\lambda_{\mathrm{WFS}} / D$. The~SR is in the lower-right corner.}
    \label{fig:images}
\end{figure}

Figure \ref{fig:images} shows the typical coronagraph images obtained with SAXO in the first row and those obtained with SAXO+ in the second row. The seeing is 0.8" and the coherence time is 3\,ms. We focus on three specific science cases, bright-1, red-1, and red-4. The system parameters (gain and frequency of the first stage, respectively $g_1$ and $f_1$, and the frequency of the second stage, $f_2$) have been optimized by minimizing our baseline criteria $C(3,5)$, the average normalized intensity between 3 and 5 $\lambda / D$ (see Sect. \ref{subsec:opt_param}).

In the bright-1 case (left) with SAXO (top) in Fig. \ref{fig:images}, the edge of the correction zone of the first stage is at 15 $\lambda / D$ from the optical axis. It is set by the number of modes controlled by the SAXO system (Sect. \ref{subsubsec:dm}). The butterfly-shaped halo is due to the AO temporal error, also known as the wind-driven halo \citep{Cantalloube_2020}. In the lower-left image, with SAXO+, the correction zone of the second stage appears inside the first one, below 10 $\lambda / D$ separation. This correction zone is smaller because the second stage DM has fewer actuators and controls fewer modes than the first stage DM. The second stage, here running at 3\,kHz, corrects most of the wind-driven halo left by the first AO loop. In the red-1 and red-4 cases (center and right images), SAXO does not provide effective turbulence correction at the wavelength of interest: There is no correction zone in the top images. The second stage of SAXO+ compensates by itself (gain of the first stage is 0.01) most of the turbulence, and the correction zone of DM2 is visible in the coronagraph images (bottom), improving the quality of the image with respect to SAXO case.

As the photon flux is decreasing in both WFSs from left to right in Fig. \ref{fig:images}, the residual intensity inside the correction zones increases, and the actual performance is reduced as expected. In particular, in the red-4 case with SAXO (upper-right image), there is no correction at all, as the SR is at 0.01. Moreover, the optimal values of first stage gain, first stage frequency, and second stage frequency decrease from left to right. As expected, we needed to slow down the loops to ensure a reasonable signal-to-noise ratio on the WFS detector.
 
\begin{figure}
    \centering
    \includegraphics[width=\hsize]{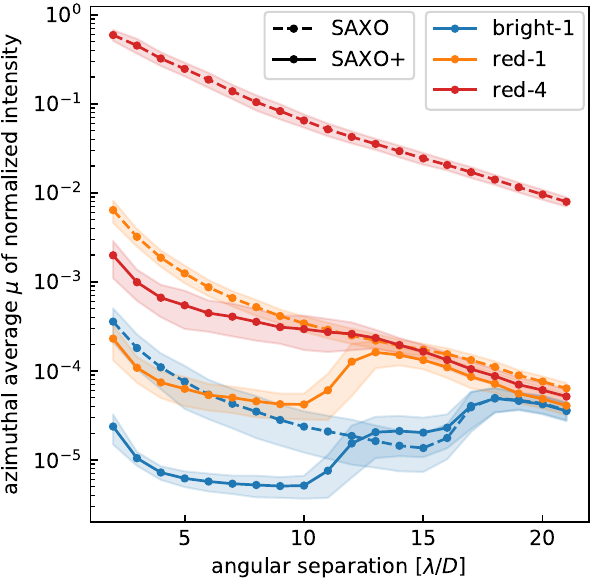}
    \caption{Azimuthal average $\mu$ of the normalized intensity of Fig. \ref{fig:images} images. The semitransparent areas represent the standard deviation of the normalized intensity. Conditions: $s = 0.8$", $\tau_0 = 3$ ms. PWFS modulation radius: 3 $\lambda_{\mathrm{WFS}} / D$. Imaging wavelength: $\lambda = 1.67$ \textmu m.}
    \label{fig:intensity}
\end{figure}

We plot in Fig.~\ref{fig:intensity} the azimuthal average $\mu$ of the normalized intensity versus the angular separation on the x-axis (see Eq.~\ref{eq:metric}). The semitransparent areas represent the azimuthal standard deviation of the normalized intensity. In our simulation, only the wind-driven halo creates asymmetry in the coronagraph images. Hence, the wider the semitransparent area in Fig.~\ref{fig:intensity}, the stronger the wind-driven halo.

For the bright case (blue curves in Fig. \ref{fig:intensity}), from 2 to 6 $\lambda / D$, the azimuthal average $\mu$ obtained with SAXO+ is lowered by a factor of ten compared to SAXO. For instance, at 5 $\lambda / D$, we obtained a $7 \cdot 10^{-5}$ normalized intensity with SAXO+. Moreover, the standard deviation (linked to the wind-driven halo, i.e., bandwidth residuals) is also reduced, by a factor two in the SAXO+ case. At a 10 $\lambda / D$ separation, we reached the edge of the second stage correction zone, with a normalized intensity of $5 \cdot 10^{-6}$ for SAXO+, improved by a factor four compared to SAXO. Further away, from $15\,\lambda / D$, the SAXO and the SAXO+ system cannot correct the turbulence, and the intensity level is dominated by the turbulence level. The correction zone cutoffs at 15 $\lambda / D$ and 10 $\lambda / D$ for the first and second stage are reduced compared to the theoretical DM geometry since we used only 800 and 400 modes, respectively (60\% and 75\% of the total number of actuators). Between $10\,\lambda/D$ and $15\,\lambda/D$, we noticed an intermediate correction zone, which is as expected. The zone is where the first stage can produce a correction, while the second stage cannot due to its limited number of actuators. It should be noted that if the frequency and the gain of the first stage are the same in the SAXO and the SAXO+ simulations, we find the SAXO and SAXO+ curves superimposed in Fig.\,\ref{fig:intensity} (between 13 and 15 $\lambda / D$). In our case the curves are not perfectly superimposed due to differently tuned AO parameters. The gain of the first stage and the frequencies of both stages are optimized for each curve to reach the best performance between~$3\,\lambda/D$ and $5\,\lambda/D$ (see Sect.~\ref{subsec:opt_param}). Improving the performance in this region comes with a loss of performance between 10 and 15 $\lambda / D$.

For the red-1 (orange) and red-4 (red) cases in Fig.\,\ref{fig:intensity}, there is a loss of performance. This is expected because there are fewer photons on the WFS and the loops have to be slowed down. Nevertheless, SAXO+ always performs better than SAXO by a factor of ten to 100 (red-4).

\subsection{First stage gain}
\label{subsec:1st_gain}

The gain of the first stage impacts the first stage residuals and, as a result, the SAXO+ performance. The overshoot of the first stage transfer function lies inside the bandwidth of the second stage transfer function. If the first stage gain is too high, it lowers the second stage correction, especially in the low photon flux regime on the SH. \citet{Cerpa-Urra_2022} showed that for a two-stage AO system in low flux conditions on the first stage, the optimal first stage gain is lower for the two-stage system than for the individual first stage. We confirmed this behavior, comparing the optimal gain of the first stage for SAXO to that of SAXO+.

\begin{figure}
    \centering
    \includegraphics[width=\hsize]{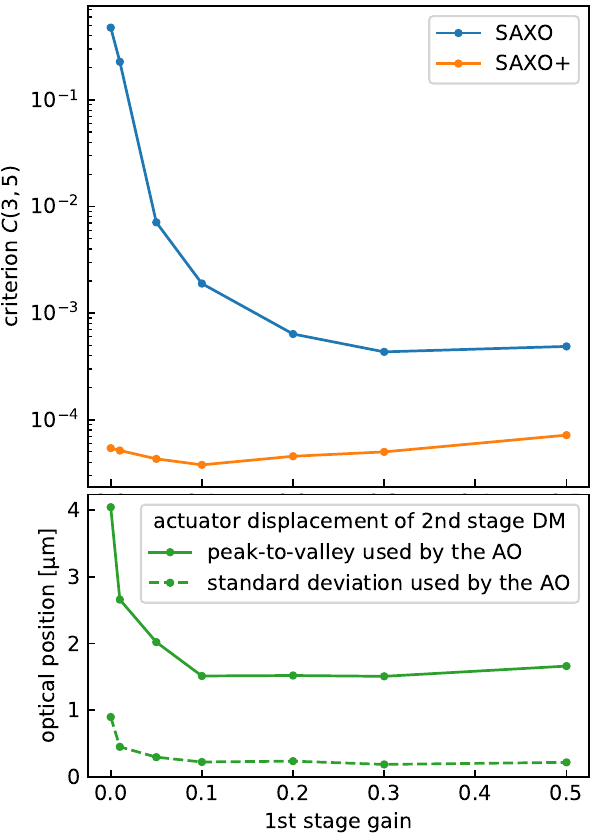}
    \caption{Impact of the first stage gain on the SAXO+ performance and the second stage DM. Upper graph: Criteria C(3, 5) (averaged normalized intensity between 3 and 5 $\lambda / D$ angular distance) versus the first stage gain. Lower graph: Maximum actuator displacement during the exposure versus the first stage gain. Science case: bright-3, seeing = 0.8", and $\tau_0 = 3$ ms. PWFS modulation radius: 3 $\lambda_{\mathrm{WFS}} / D$. Imaging wavelength: $\lambda = 1.67$ \textmu m.}
    \label{fig:gain_stroke}
\end{figure}

In the upper plot of Fig. \ref{fig:gain_stroke}, we represent the criterion $C(3, 5)$ -- the average normalized intensity between 3 and 5 $\lambda / D$ -- as a function of the first stage gain, with SAXO alone in blue and with SAXO+ in orange. The science case is bright-3, the seeing is 0.8", and the coherence time is $3\,$ms. The PWFS modulation radius is 3 $\lambda_{\mathrm{WFS}} / D$. For both the SAXO and SAXO+ curves, there is an optimal first stage gain for which the performance criterion is minimal. In these specific observing conditions, the best performance of SAXO is achieved with a 0.3 optimal gain, while the best performance of SAXO+ is reached with a lower optimal gain of 0.1 on the first stage. This behavior confirms the results of \citet{Cerpa-Urra_2022}.

Nevertheless, reducing the first stage gain increases the amount of turbulence that the second stage has to correct. The specifications of Boston Micromachines on their DM indicates an optical stroke of 11 \textmu m. Part of this stroke is used to flatten the DM, correct for NCPA, and other additional calibrations. Hence, for the second stage DM, the stroke might not be sufficient to deal with the first stage residuals. In our simulations, we consider no stroke limitations, and the second stage DM has an infinite stroke. However, we recorded the actuator displacement during each simulation.

In the lower part of Fig. \ref{fig:gain_stroke}, for each first stage gain, we have plotted the temporal maximum of the spatial peak-to-valley displacement over the DM actuators of the second stage (full line) as well as the temporal maximum of the spatial standard deviation (dashed line). Between 0.1 and 0.5, the maximum actuator displacement is at a constant level of 1.5 \textmu m, and the maximum standard deviation displacement is 0.3 \textmu m. When the first stage gain drops below 0.1, the maximum peak-to-valley displacement increases and reaches 2.7 \textmu m for a 0.01 gain and 4 \textmu m when the first stage is deactivated (gain is zero). For the considered science case (bright-3) and turbulence conditions (0.8" seeing and 3\,ms coherence time), setting the first stage gain at its optimal value of 0.1 seems to be reasonable, as 1.5 \textmu m of actuator displacement is used while the stroke of the DM is 11 \textmu m.

We extended this analysis to all SAXO+ science cases and turbulence conditions. We obtained a similar behavior for all cases, and only the values of the optimal gain and the peak-to-valley displacement changed (Sect. \ref{subsec:opt_param}).

\subsection{Second stage frequency}
\label{subsec:2nd_freq}

The maximum frequency of the second stage is one of the major system trade-offs. In high flux conditions, a higher frequency means a lower stellar intensity in coronagraph images but harsher constraints on the real-time system (e.g., pyramid modulation mirror, RTC, electronics). In this section, we study the performance as a function of the second stage frequency.

\begin{figure*}
    \sidecaption
    \centering
    \includegraphics[width=12cm]{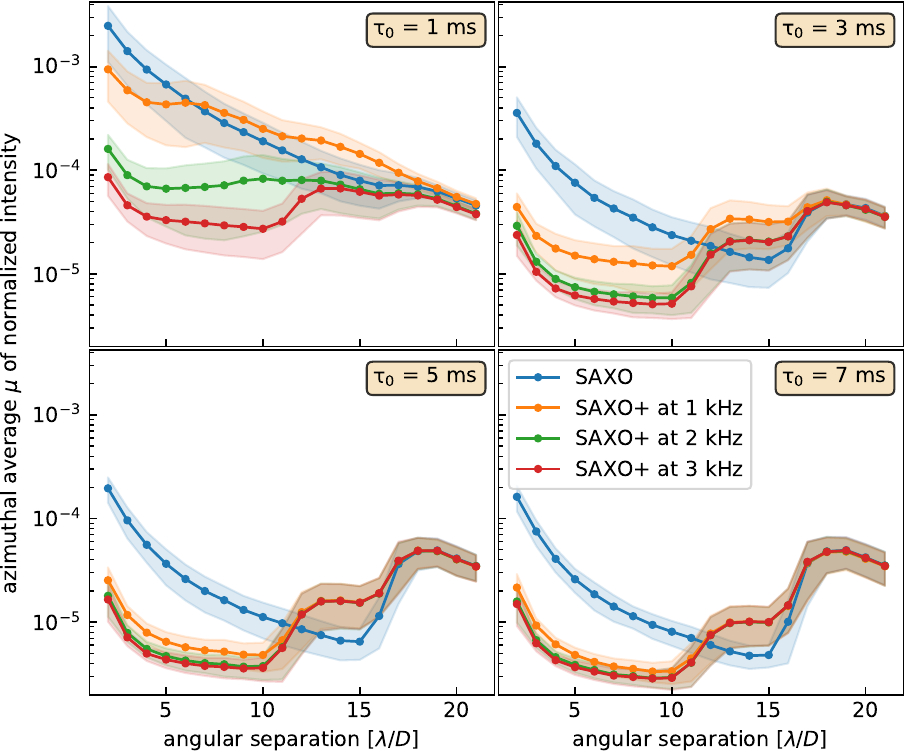}
    \caption{Azimuthal average $\mu$ of the normalized intensity for different frequencies of the second stage. Conditions: seeing = 0.8", science case = bright-1. PWFS modulation radius: 3 $\lambda_{\mathrm{WFS}} / D$. Imaging wavelength: $\lambda = 1.67$ \textmu m.}
    \label{fig:freq2}
\end{figure*}

Figure \ref{fig:freq2} shows the azimuthal average $\mu$ of the normalized intensity versus the angular separation for SAXO (in blue) and for SAXO+ with various frequencies of the second stage (1 kHz in orange, 2 kHz in green, and 3 kHz in red) and four coherence times (1, 3, 5, 7 ms). The science case is bright-1, namely high flux conditions, and the seeing is 0.8". The PWFS modulation radius is 3 $\lambda_{\mathrm{WFS}} / D$.

For all cases ($\tau_0$ and $f_2$) but one that we discuss below ($\tau_0 = 1$ ms and $f_2 = 1$ kHz), SAXO+ improves the coronagraph performance between 3 and 5 $\lambda / D$ by a factor of ten or five. In all cases, the SAXO+ performance improves as $f_2$ increases. And the azimuthal average $\mu$ of the normalized intensity is always smaller at 3 kHz than at 2 kHz. The performance gain between the different operating frequencies narrows as the coherence time increases, which is expected because the bandwidth error becomes less and less dominant.

If the second stage runs at 1 kHz and $\tau_0 =1\,$ ms, SAXO+ does not improve the correction of SAXO beyond 6 $\lambda / D$, as the blue and orange curves are superimposed in Fig.\,\ref{fig:freq2} in the upper-left graph. This happens because the CLOSE algorithm lowers the modal gains on the highest order modes for the SAXO+ system to compensate for the low signal-to-noise ratio on the PWFS. To overcome this phenomenon, we could use a modal integrator on the first stage and increase the gain on high-order modes above 6 $\lambda / D$. Not only in this specific case but more generally, increasing the first stage gain for high-order modes above 10 $\lambda / D$ (modes not corrected by the second stage) might improve the SAXO+ performance, compared to our current study. Indeed, it will decrease the normalized intensity between 10 and 15 $\lambda / D$, as the first stage gain is currently suboptimal for a SAXO-alone system. Moreover, it will improve the PSF quality at the top of the pyramid WFS and reduce aliasing effects.

\begin{figure}
    \centering
    \includegraphics[width=\hsize]{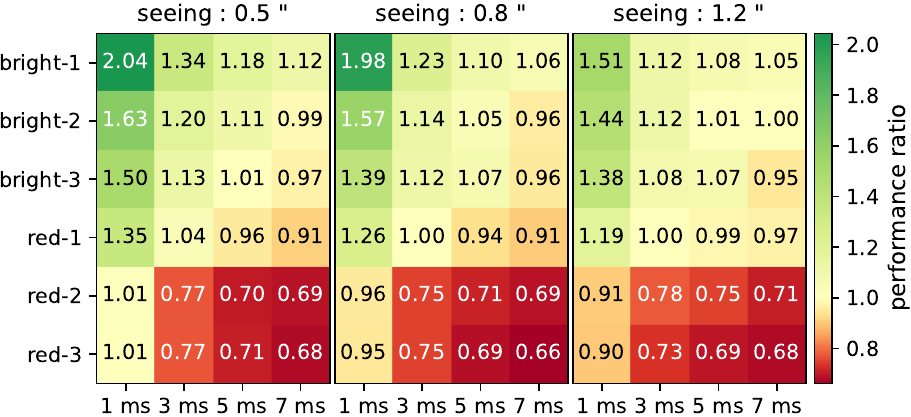}
    \caption{Ratio between performance at 2 kHz and 3 kHz. Here the performance criteria is the average intensity between 3 and 5 $\lambda / D$. Green means 3 kHz performs better than 2 kHz, while red means 2 kHz performs better than 3 kHz.}
    \label{fig:ratio}
\end{figure}

In Fig. \ref{fig:ratio}, we give the average intensity between 3 and 5 $\lambda / D$ with a 2 kHz second stage divided by the same quantity for the 3 kHz case. A ratio greater than one (green boxes) means that 3 kHz performs better than 2 kHz, while a ratio lower than one (red boxes) means that 2 kHz performs better than 3 kHz. The coherence time varies along the rows (values at the bottom), the science cases vary along the columns, and each table corresponds to a given seeing (values at the top). For all cases, the ratio of normalized intensities decreases as the coherence time increases, as expected and as we already commented about for Fig. \ref{fig:freq2}: the longer the coherence time, the smaller the correction loop frequency can be.

Moreover, the ratio of normalized intensities decreases with the star brightness. With a 0.8" seeing and a 3 ms coherence time, the ratio is 1.23 in the bright-1 case, and it decreases to 0.75 in the red-3 case. In most of the red cases, 3 kHz is too fast, which means that there are not enough photons on the WFS to get a better correction by increasing the speed of the loop.

Except for the lowest coherence time of 1 ms, the ratio of performance does not exceed 1.34 (in the bright-1 case with 0.5" seeing and $\tau_0$ = 3 ms). Given the constraints for 3 kHz (RTC, pyramid modulation, electronics), this improvement of performance may not be significant enough to justify an increase in the maximum second stage frequency from 2 kHz to 3 kHz.

\subsection{Optimal parameters}
\label{subsec:opt_param}

\begin{figure*}
    \centering
    \includegraphics[width=\textwidth]{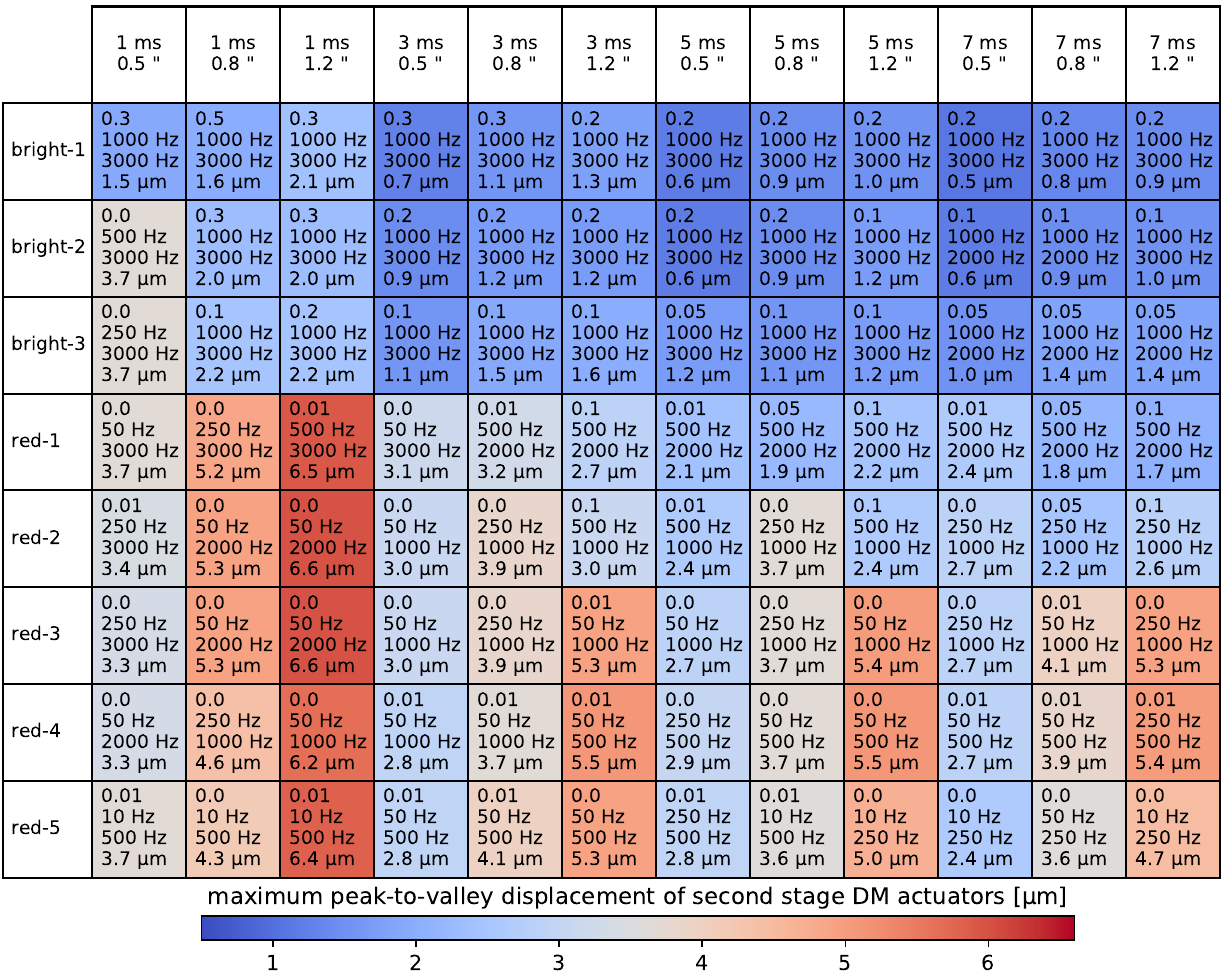}
    \caption{SAXO+ optimal parameters and maximum peak-to-valley displacement of second stage DM actuators with respect to the science cases (rows) and the turbulence conditions (columns). Each box contains (in this order) the first stage gain, the first stage frequency, the second stage frequency, and the maximum peak-to-valley displacement.}
    \label{fig:optimals2}
\end{figure*}

This section synthesizes the optimization of the system parameters for all observing conditions with a fixed 3 $\lambda_{\mathrm{WFS}} / D$ PWFS modulation radius in Fig.\,\ref{fig:optimals2}. We first set the observing conditions, for instance a 3 ms coherence time, a 0.8" seeing, and the bright-3 science case. We ran simulations for every combination of first stage gain $g_1$ (Table \ref{param_1}) and first and second stage frequencies (Table \ref{param_2}), respectively $f_1$ and $f_2$. Then we chose the simulation with the triplet of parameters $g_1$, $f_1$, and $f_2$ that minimizes the criteria $C(3, 5)$. For a bright-3 science case, a 3 ms coherence time, and a 0.8" seeing (third row and fifth column in Fig.\,\ref{fig:optimals2}), the optimal parameters are $g_1 =$~0.1, $f_1 =$~1 kHz, and $f_2 =$~ 3 kHz. The maximum peak-to-valley displacement of an actuator of the second stage DM is 1.6 \textmu m.

This optimization was performed again for each combination of science case, coherence time, and seeing, which built the rest of the table of Fig. \ref{fig:optimals2}. Each box contains, from top to bottom, the optimal first stage gain, the optimal first and second stage frequencies, and the peak-to-valley displacement. The color scale refers to the peak-to-valley displacement.

The optimal second stage frequency decreases from the upper-left corner to the bottom-right corner. This is expected because the optimal second stage frequency decreases for fainter photon flux on the PWFS and as the coherence time of the atmosphere increases. Another consistent observation is the decrease of the optimal first stage gain with the photon flux on the SH WFS. Indeed, decreasing the first stage gain mitigates the propagation of the noise and the aliasing in the first stage residuals. For instance, with a 3 ms coherence time and a 0.8" seeing, the optimal first stage gain is 0.3 for the bright-1 case, and it reduces between zero and 0.1 for every red case. A gain of zero is equivalent to an open loop for the first stage.

The maximum peak-to-valley displacement of the second stage DM actuators is also represented with the color scale. Blue cases are the lowest peak-to-valley displacement, and red cases are the highest. As wavefront correction by the first stage degrades, from top to bottom, the maximum peak-to-valley displacement increases because the second stage has to correct for more and more aberrations. In the worst case ($\tau_0 = 1$~ms, seeing = 1.2", red-2, and red-3 cases), the maximum displacement reaches~6.6\,\textmu m. However, for red-1 and red-2 cases, we found that the actuator displacement can be reduced by about 1 \textmu m by increasing the first stage gain to 0.1, and the loss in performance is by no more than a factor of 1.2 for all turbulence conditions.

\begin{figure*}
    \centering
    \includegraphics[width=\textwidth]{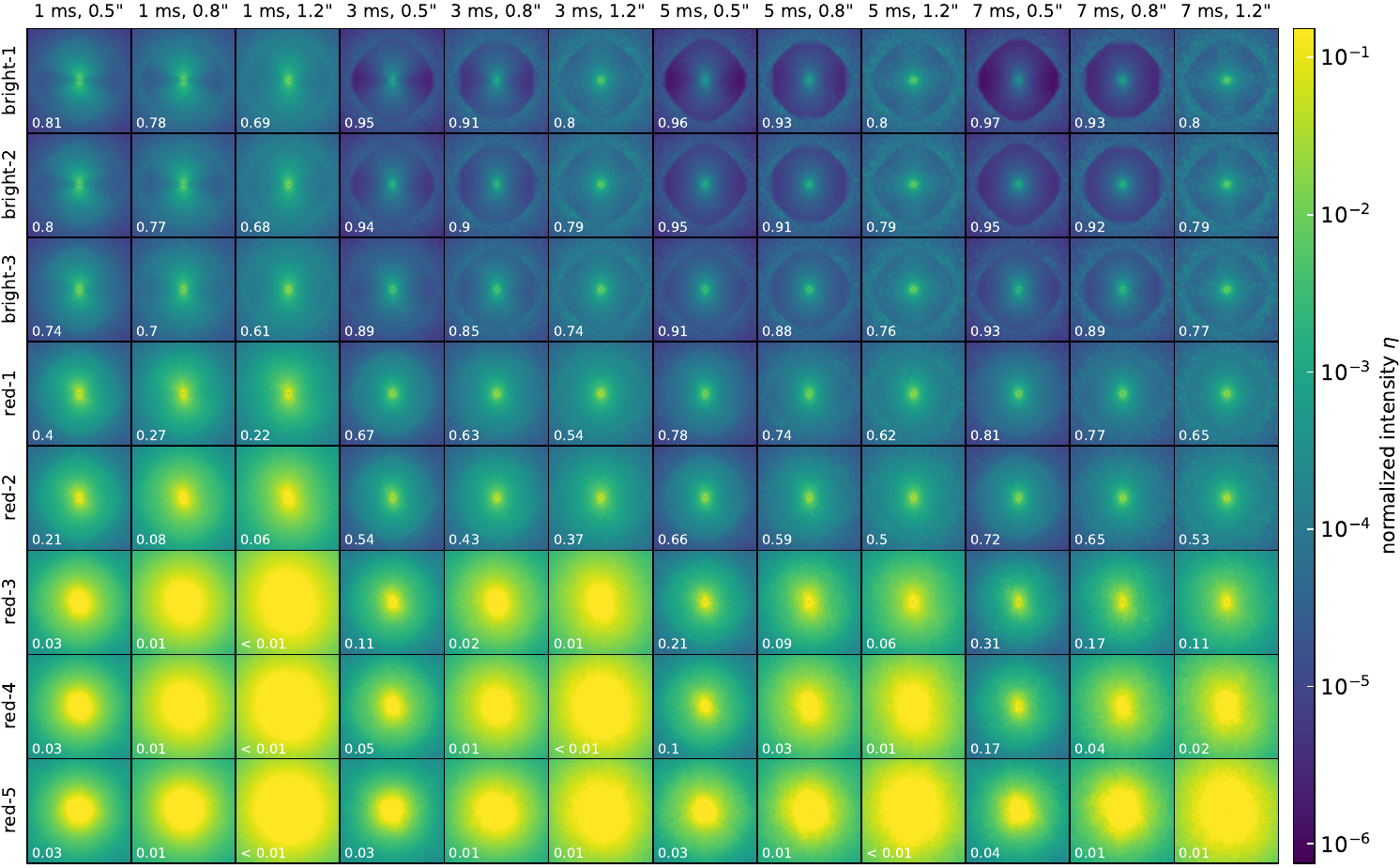}
    \caption{Coronagraph images after SAXO for all science cases (rows) and atmospheric conditions (columns). Imaging wavelength: $\lambda = 1.67$ \textmu m.}
    \label{fig:images96saxo}
\end{figure*}

\begin{figure*}
    \centering
    \includegraphics[width=\textwidth]{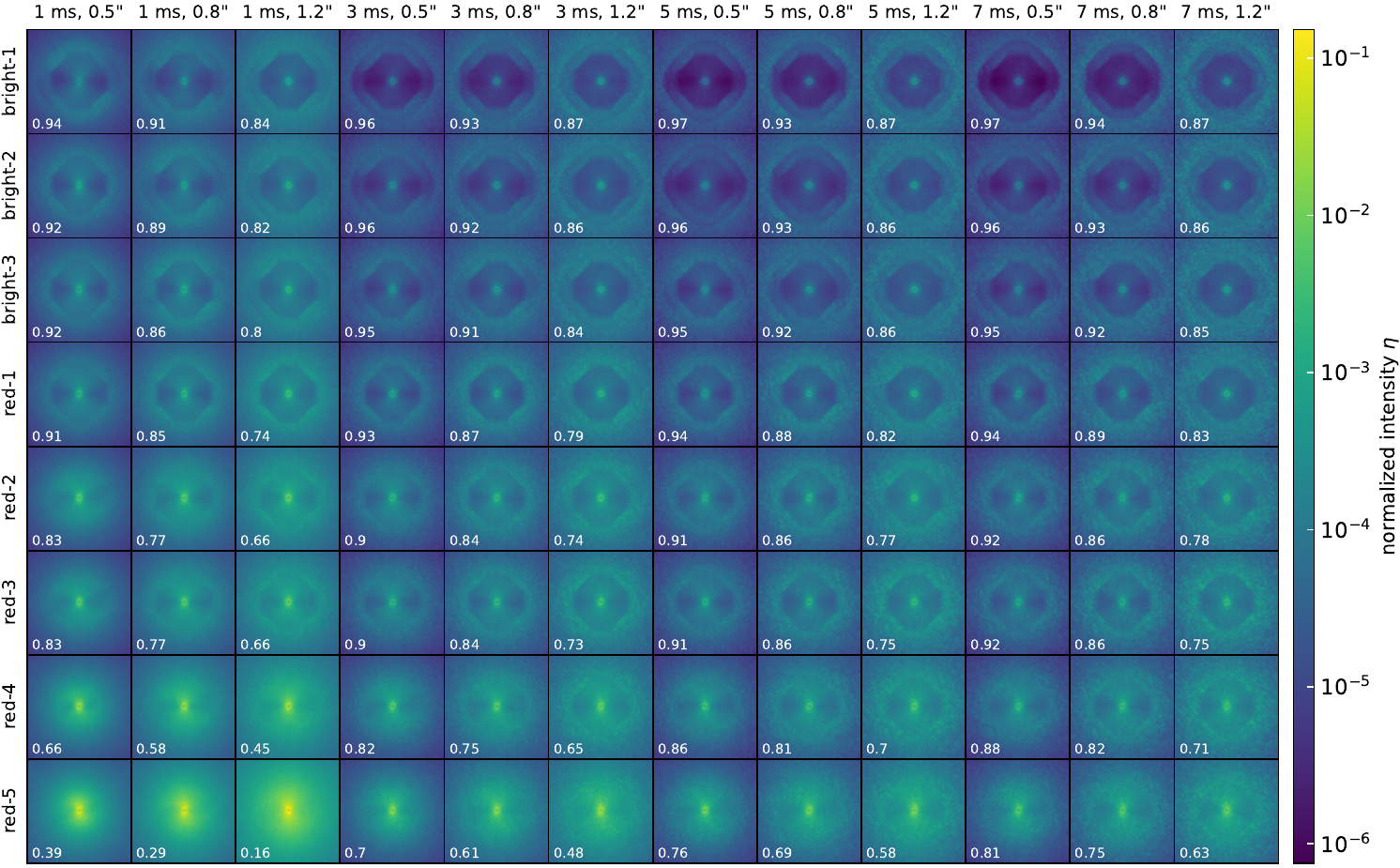}
    \caption{Coronagraph images after SAXO+ for all science cases (rows) and atmospheric conditions (columns). Imaging wavelength: $\lambda = 1.67$ \textmu m.}
    \label{fig:images96}
\end{figure*}

In Fig. \ref{fig:images96saxo} and \ref{fig:images96}, we show the coronagraph images obtained with the optimal parameters for SAXO and SAXO+, respectively, for all science cases (rows) and observing conditions (columns). The optimal parameters of SAXO alone are presented in Appendix A. They are not the same as the first stage in SAXO+. For example, the optimal first stage gain can be different in SAXO and SAXO+ (Sect.~\ref{subsec:1st_gain}). The SR is written in the lower-left corner of each image. In the bright-1 to bright-3 cases, SAXO+ creates a deeper correction zone inside the SAXO one. For the red cases, SAXO achieves a poor correction (SR lower than 0.7) or no correction at all (SR $\sim$ 0.1 and no correction zone). That is why the optimal first stage gain for SAXO+ is mostly zero or 0.01 in the red cases (see Fig. \ref{fig:optimals2}). Nevertheless, with SAXO+, the second stage correction zone is present in the coronagraph images, and SAXO+ performs better than SAXO in all science cases and turbulence conditions.

\subsection{Comparison with real SAXO performance}
\label{subsec:compar_real_instru}
In this section, we show that the absolute level of our simulated SAXO performance is in good agreement with the measured on-sky performance. Several studies have reported the performance of SAXO. However, the instrument has evolved since 2014~\citep[e.g., low wind effect limitation, ][]{Milli_2018_lowwindeffect} and so have its performance criteria~\citep[e.g., discrepencies between RTC-Strehl and measured Strehl, ][]{Milli_2017}. We therefore chose to compare our simulation only with the latest on-sky SAXO performance published \citep{Jones_2022, courtney-barrer2023_EmpiricalContrastModel}.

For the red-1 (G~mag~$= 11.9$) and red-2 (G~mag~$ = 12.8$) cases, \citet{Jones_2022} measured an on-sky SR at $0.52 +/- 0.08$ and $0.43 +/- 0.12$, respectively, in the 30\% best condition category (corresponding to a seeing of \,$< 0.8"$ and $\tau_0> 4.1\,$ms). These observing conditions are between the $3$ and~$5\,$ms cases for $0.8''$ seeing in our study (probably closer to the $3\,$ms case, as the~SR performance drops fast for $\tau_0<5\,$ms). The on-sky performance is in good agreement with our simulated~SR (Fig. \ref{fig:images96saxo}) between 0.63 and 0.74 (slightly optimistic) for red-1 and between 0.43 and 0.59 (excellent agreement) for red-2. The fainter cases studied in \citet{Jones_2022} are G~mag~$= 13.9$; therefore, we can only put an upper limit for the performance for red-3 (G~mag~$= 14.5$) at~SR$< 0.05$. This is in agreement with our simulations that lie between 0.02 and 0.09.\\

The use of a perfect coronagraph (Sect.~\ref{subsec:coro}) and the absence of NCPA in our simulations likely lead to an overestimation of the SPHERE normalized intensity performance in the best observing conditions. For these reasons, we only compare our simulation to the worst atmospheric condition cases in \citet{courtney-barrer2023_EmpiricalContrastModel}, for which the performance is more likely limited by AO residuals only. The bright-2 (G~mag~$=7.6$) and bright-3 (G~mag~$=9.6$) cases in our study respectively correspond to mid targets ($5<$\,G~mag\,$<9$) and faint targets (G~mag\,$>9$) in \citet{courtney-barrer2023_EmpiricalContrastModel}. From their Fig.~8, we found a difference of less than~0.5 magnitude between their on-sky measurements in the worst turbulence case (best 85\% category) and our simulation for the $3$~ms case for $1.2''$ seeing for both bright-2 and bright-3 cases.

\subsection{Modulation radius of the PWFS}
\label{subsec:modu}

The linearity range and the robustness of the PWFS increases with the modulation radius but at the cost of sensitivity. Moreover, in the SAXO+ system with a 3 kHz second stage, the modulation mirror can hardly draw a modulation larger than 3 $\lambda_{\mathrm{WFS}} / D$. In the following sections, we study the impact of PWFS modulation on the coronagraph performance.

We performed simulations with various values of modulation radius, science cases, seeing, and second stage gain $g_2$. The explored parameter space is presented in Table \ref{table:modu}. For each science case, the first and second stage frequencies and the first stage gain are fixed to the optimal values obtained in paragraph \ref{subsec:opt_param} for the 3 $\lambda_{\mathrm{WFS}} / D$ modulation (Figure \ref{fig:optimals2}):
\begin{itemize}
    \item bright-1 : $g_1$ = 0.3, $f_1$ = 1000 Hz, $f_2$ = 3000 Hz;
    \item red-1 : $g_1$ = 0.1, $f_1$ = 500 Hz, $f_2$ = 2000 Hz;
    \item red-4 : $g_1$ = 0.01, $f_1$ = 50 Hz, $f_2$ = 1000 Hz.
\end{itemize}
The coherence time is fixed at $\tau_0\,=\,3\,$ms. We also performed the simulations with and without the CLOSE optimization on the second stage integrator in order to evaluate the impact of CLOSE on the overall performance. When CLOSE is disabled, the vector of modal gains $\vec{g}_m$ (Eq. \ref{eq:integrator2}) is constant and equal to one.

\begin{table}
    \renewcommand{\arraystretch}{1.12}
    \centering
    \caption{Explored parameters for simulations about modulation radius.}
    \begin{tabular}{ll}
        modulation radius & no modulation, 1, 2, 3, 5, 10 $\lambda_{\mathrm{WFS}} / D$ \\
        science cases & bright-1, red-1 and red-4 \\
        seeing & 0.8", 1.2" \\
        $2^{\mathrm{nd}}$ stage gain $g_2$ & 0.1, 0.2, 0.4, 0.6, 0.8, 1 \\
        modal gains $\vec{g}_m$ & optimized by CLOSE or fixed to 1
    \end{tabular}
    \label{table:modu}
\end{table}

\subsubsection{CLOSE versus manually optimized scalar gain}
\label{subsubsec:close}

\begin{figure}
    \centering
    \includegraphics[width=\hsize]{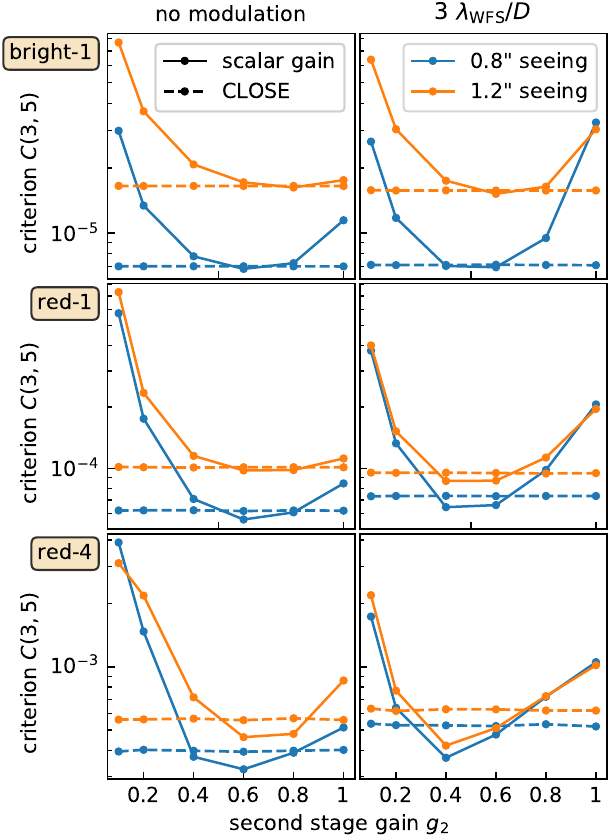}
    \caption{Criterion of performance $C(3, 5)$ versus second stage gain for three science cases (bright-1, red-1, and red-4) and two seeing conditions (0.8" in blue and 1.2" in orange). We compare the CLOSE optimization (dashed line) and the scalar gain case (without CLOSE, full line). Imaging wavelength: $\lambda = 1.67$ \textmu m.}
    \label{fig:gain2}
\end{figure}

In this section, we study the impact of the loop gain $g_2$ on the coronagraph performance, with the CLOSE optimization and in the scalar gain case (when CLOSE is disabled). Figure \ref{fig:gain2} shows the average of the normalized intensity between 3 and 5 $\lambda / D$ -- performance criterion $C(3, 5)$ -- versus the second stage $g_2$. The dashed lines are simulations with the CLOSE optimization, and the solid lines are those with a scalar gain. Blue curves correspond to 0.8" seeing and orange curves to 1.2" seeing. In the figure, we compare a no modulation case on the left with a 3 $\lambda_{\mathrm{WFS}} / D$ modulation radius on the right.

In the scalar gain case, when CLOSE is disabled (solid lines), there is an optimal gain of the second stage integrator that minimizes the $C(3, 5)$ criterion. For instance, in the bright-1 case, the optimal gain is 0.6 for a 0.8" seeing and 0.8 for a 1.2" seeing. With the CLOSE optimization, the dashed lines are flat, which is expected because CLOSE does a temporal optimization of each modal gain in the vector $\vec{g}_m$. If the scalar gain $g_2$ is below its optimal value, CLOSE will compensate for it by increasing the modal gains $\vec{g}_m$, and vice versa.

In the bright-1 case, the minimum of the scalar gain curve is equal to the performance achieved with CLOSE. For instance, with a 0.8" seeing and a 3 $\lambda_{\mathrm{WFS}} / D$ modulation radius, the performance criteria $C(3, 5)$ is $7 \cdot 10^{-6}$ for the dashed and solid lines. In the red-1 and red-4 cases, the minima of the scalar gain curves are slightly lower than the performance we obtained with the CLOSE optimization. The most obvious case is red-4 with a 3 $\lambda_{\mathrm{WFS}} / D$ modulation radius, where there is a factor of 1.5 difference between the best scalar gain performance and the CLOSE performance. The CLOSE algorithm is less efficient on fainter targets because the signal-to-noise ratio of the PWFS measurements is lower compared to bright targets, which complicates the transfer function retrieval.

Finally, we found that a manual optimization of the scalar gain provides a performance similar to the automatic CLOSE optimization. The factor of 1.5 difference between the best scalar gain performance and the CLOSE performance is indeed small. As atmospheric conditions (seeing, coherence time) often vary during observations, the CLOSE algorithm is more robust than a manually optimized scalar gain. Moreover, on sky, the curve performance versus scalar gain (used to find the optimal scalar gain) are noisy and hardly measurable.

However, there is no improvement from using a modal optimization compared to an optimized scalar gain. The dependency of the PWFS optical gains with the spatial order of the mode is not significant in the SAXO+ system. This is expected because the PWFS wavelength is in the near-infrared (1.2 \textmu m), and the SR ratio at the top of the pyramid is quite high (60 \% in the red-4 case with 0.8" seeing). Thus, the PWFS works in a sufficiently linear regime to avoid the need of dealing with the spatial dependency of the optical gains.

\subsubsection{Impact of modulation}

\begin{figure}
    \centering
    \includegraphics[width=0.9\hsize]{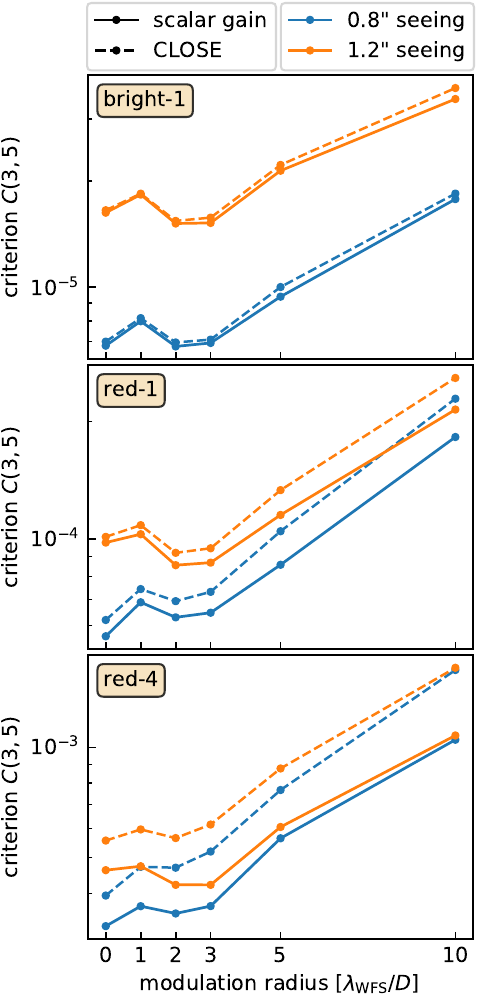}
    \caption{Criterion of performance $C(3, 5)$ versus pyramid modulation radius for three science cases (bright-1, red-1, and red-4) and two seeing conditions (0.8" in blue and 1.2" in orange). We compare the CLOSE optimization (dashed line) and a manually optimized scalar gain (without CLOSE, full line). Imaging wavelength: $\lambda = 1.67$ \textmu m.}
    \label{fig:rmod}
\end{figure}

In Fig. \ref{fig:rmod}, we plot the average of the normalized intensity between 3 and 5 $\lambda / D$ versus the modulation radius of the PWFS using either a manually optimized scalar gain (full line) or the CLOSE optimization (dashed line). A 0 $\lambda_{\mathrm{WFS}} / D$ modulation radius means that there is no modulation at all. In the scalar gain case, we optimized the second stage gain as described  in paragraph \ref{subsubsec:close} (see Fig. \ref{fig:gain2}).

First, we analyzed the radii above 1 $\lambda_{\mathrm{WFS}} / D$, and we did not consider the nonmodulated pyramid. In all simulated cases, the optimal modulation radius is 2 $\lambda_{\mathrm{WFS}} / D$. As the modulation radius increases, the performance degrades up to a factor three at 10 $\lambda_{\mathrm{WFS}} / D$. Such behavior is expected because the sensitivity of the PWFS decreases as the modulation radius increases. The performance at 1 $\lambda_{\mathrm{WFS}} / D$ is higher than at 2 $\lambda_{\mathrm{WFS}} / D$. This can be explained by the decrease in the PWFS linearity range with the modulation radius.

In all cases, the performance of the nonmodulated pyramid is better than a 1 $\lambda_{\mathrm{WFS}} / D$ modulation radius and similar to the 2 $\lambda_{\mathrm{WFS}} / D$ case. It is possible that the gain in sensitivity (by decreasing the modulation radius from 1 to 0 $\lambda_{\mathrm{WFS}} / D$) is more significant in the final performance than the loss in linearity range. For a 0.8" seeing (blue curves), the best performance is achieved with a nonmodulated PWFS for the three target cases. But with a 1.2" seeing, the optimal modulation radius is still 2 $\lambda_{\mathrm{WFS}} / D$. The required linearity range of the PWFS widens when AO residuals increase. The performance gain of the nonmodulated pyramid is small, though, and the modulated pyramid is more robust against condition variations.

\section{Conclusions}

Currently in the design phase, SAXO+ is the planned upgrade of SAXO, the AO system of the exoplanet imager SPHERE at the VLT. The SAXO+ upgrade will consist of a second stage AO downstream of the SAXO stage. This second loop will be faster than the first and includes a near-infrared PWFS. In this work, we ran end-to-end simulations in order to dimension this system.

To assess the performance of SAXO+, we computed the perfect coronagraph image of an on-axis source. The explored parameter space (more than 3000 cases) includes science cases, turbulence conditions (seeing and coherence time), and system parameters (first stage gain, first and second stage frequency, PWFS modulation radius, PWFS modal gains optimization). For all science cases and conditions, we have shown that SAXO+ reduces the residual starlight intensity by a factor of ten inside the correction zone of the second stage, between 3 and 10 $\lambda / D$ (Fig. \ref{fig:images96saxo} and \ref{fig:images96}).

To find the optimal system parameters, we used as a performance criteria, the average normalized intensity between 3 and 5 $\lambda / D$. The optimal first stage gain is lower for SAXO+ than for SAXO alone (without the second stage). A second stage frequency of 2 kHz seems a realistic trade-off between the instrument performance and technical constraints on the real-time system (RTC, electronics, PWFS modulation). We summarize the simulation results in a table with the optimal values of the three system parameters for all simulated science cases and turbulence conditions (Fig. $\ref{fig:optimals2}$). Based on specific simulations for the PWFS, we suggest that a $2 \lambda / D$ modulation radius is a good compromise between performance and robustness against varying turbulence conditions. Finally, we find that the SAXO+ system can be optimized to fulfill the requirements provided in \citet{Boccaletti_2020}, namely observing red stars and improving the performance of the current system.

We remind that these results are valid under the assumption of two independent integrators: one with a scalar gain on the first stage and one with the CLOSE algorithm on the second stage (with CLOSE or with a scalar gain for the results about the pyramid modulation radius). Forthcoming studies with more recent and efficient control techniques might update our conclusions.

With the SAXO+ performance described in this paper, the actual performance of the instrument will often be limited by two effects that were purposefully neglected in this paper: the coronagraph diffraction pattern and the quasi-static speckles due to NCPA.  In a forthcoming paper, we will enhance this study with realistic coronagraphs and a third control loop in cascade based on focal plane wavefront sensing \citep{Potier_2022}.

\begin{acknowledgements}
This research has made use of data obtained from or tools provided by the portal exoplanet.eu of The Extrasolar Planets Encyclopaedia

\end{acknowledgements}

\bibliographystyle{aa} 
\bibliography{biblio} 

\begin{appendix}

\section{Optimal parameters of SAXO alone}

\begin{figure*}
    \centering
    \includegraphics[width=0.95\textwidth]{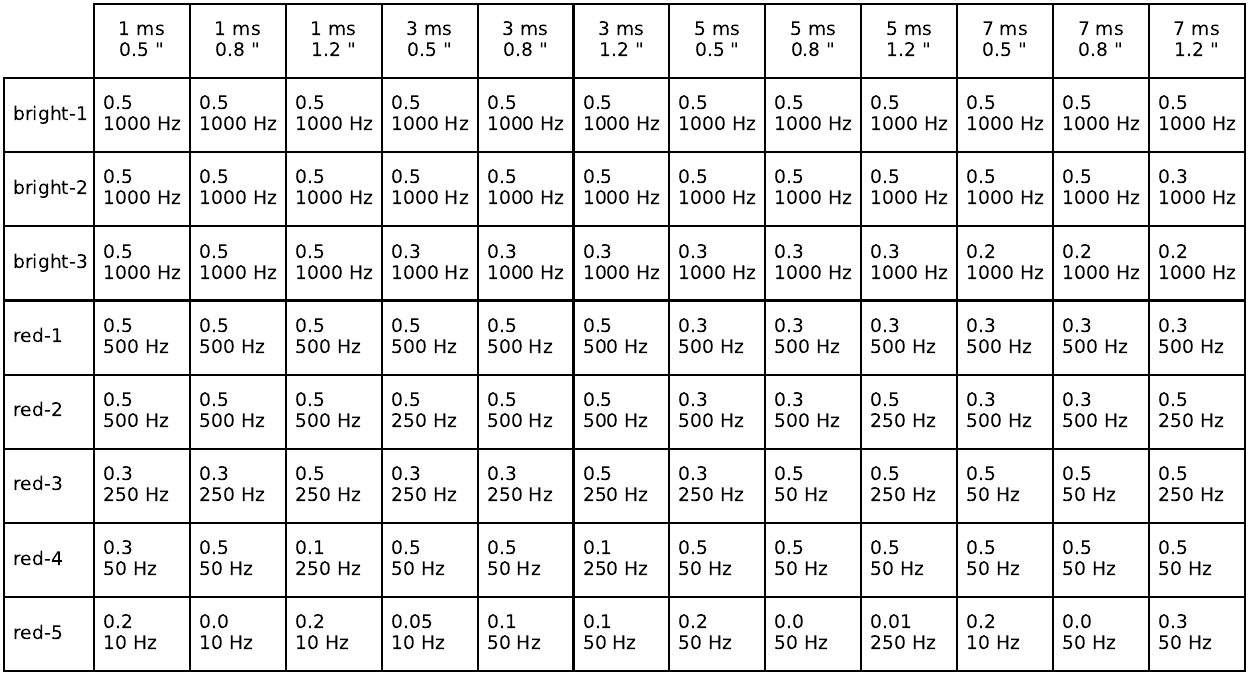}
    \caption{First stage optimal parameters with respect to the science cases and the turbulence conditions. Each box contains the optimal first stage gain and the optimal first stage frequency.}
    \label{fig:optimals1}
\end{figure*}

We show in Fig. \ref{fig:optimals1} the optimal first stage gain and frequency for SAXO alone, that is to say when the first stage runs without the second stage. As highlighted for a specific case in Sect.~\ref{subsec:1st_gain}, the optimal first stage gain is always lower for SAXO+ than for SAXO. For instance, in the bright-1 case, the optimal gain is~0.5 for SAXO while the optimal first stage gain for SAXO+ is~0.3 or~0.2 depending on the turbulence conditions (except 0.8" seeing and $\tau_0$ = 1 ms).

\section{Azimuthal average $\mu$ of normalized intensity for every science cases and turbulence conditions.}

\begin{figure}
    \centering
    \includegraphics[width=0.95\hsize]{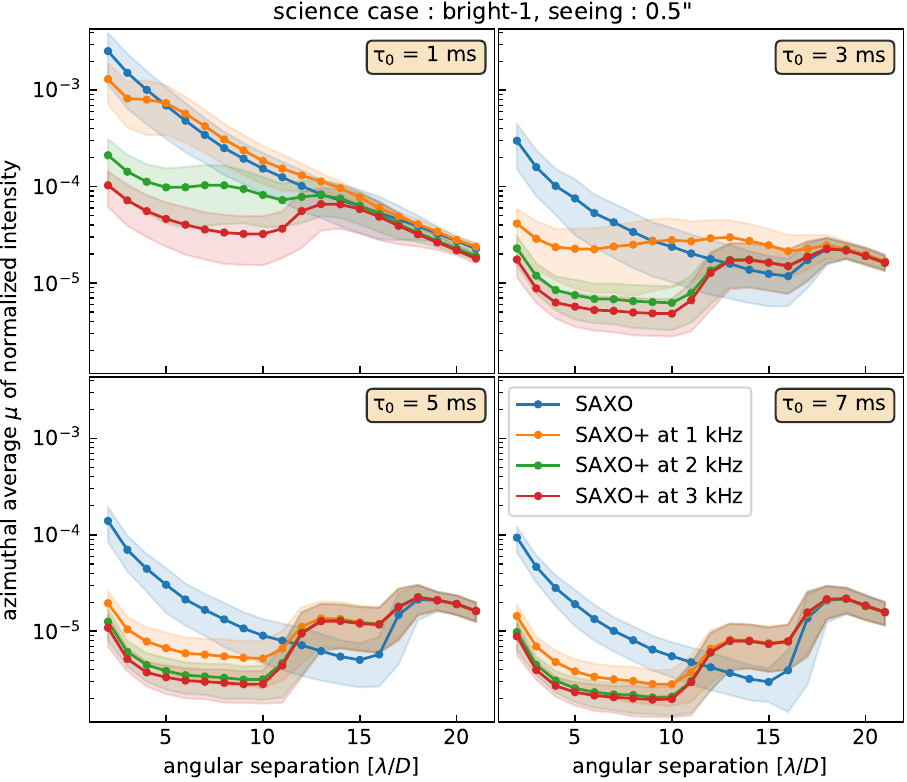}
    \caption{Bright-1, seeing = 0.5".}
    \label{fig:freq2_bright1_seeing_0.5}
\end{figure}

\begin{figure}
    \centering
    \includegraphics[width=0.95\hsize]{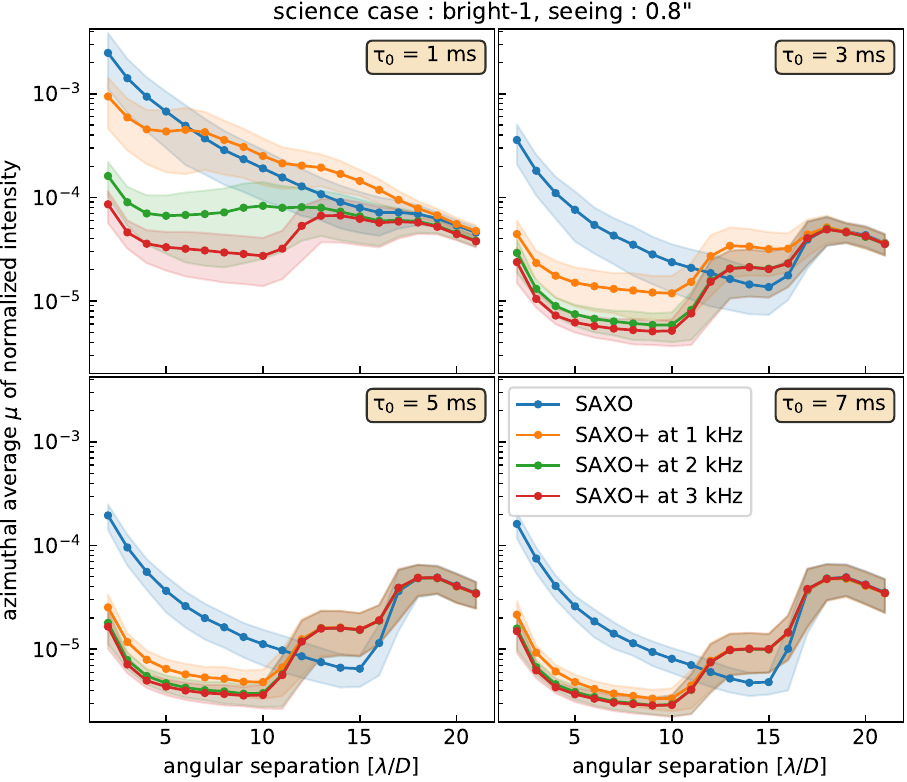}
    \caption{Bright-1, seeing = 0.8".}
    \label{fig:freq2_bright1_seeing_0.8}
\end{figure}

\begin{figure}
    \centering
    \includegraphics[width=0.95\hsize]{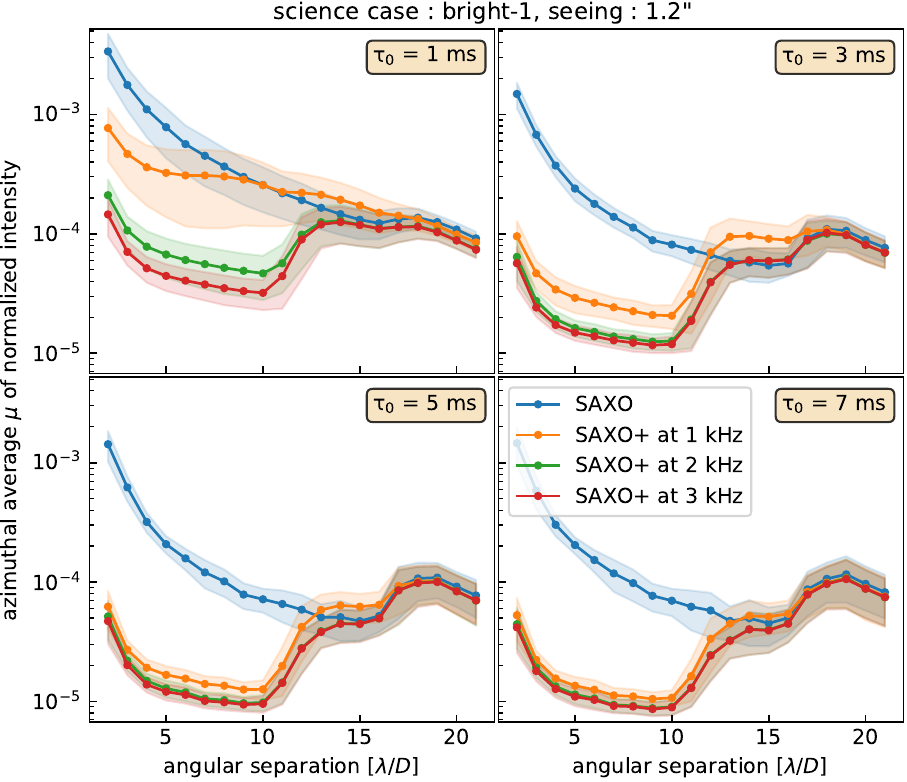}
    \caption{Bright-1, seeing = 1.2".}
    \label{fig:freq2_bright1_seeing_1.2}
\end{figure}

\begin{figure}
    \centering
    \includegraphics[width=0.95\hsize]{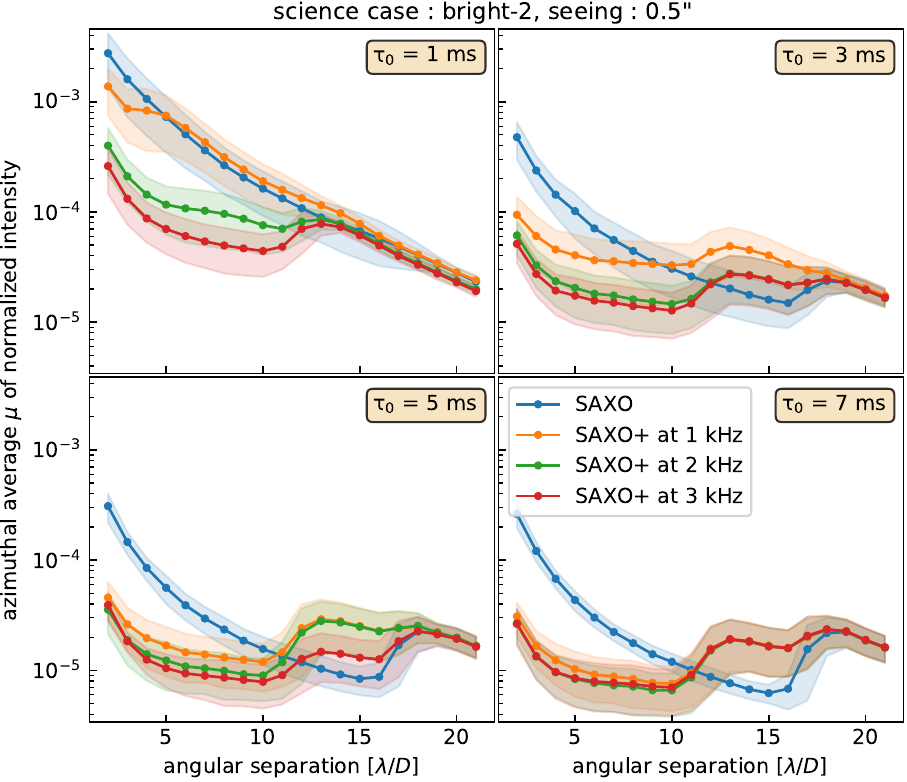}
    \caption{Bright-2, seeing = 0.5".}
    \label{fig:freq2_bright2_seeing_0.5}
\end{figure}

\begin{figure}
    \centering
    \includegraphics[width=0.95\hsize]{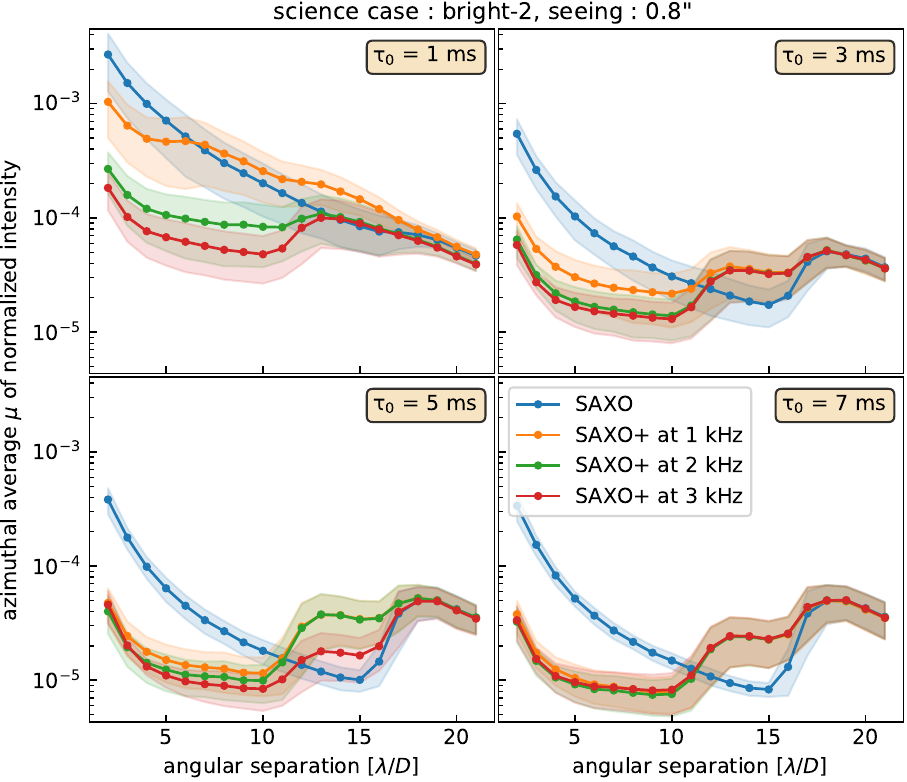}
    \caption{Bright-2, seeing = 0.8".}
    \label{fig:freq2_bright2_seeing_0.8}
\end{figure}

\begin{figure}
    \centering
    \includegraphics[width=0.95\hsize]{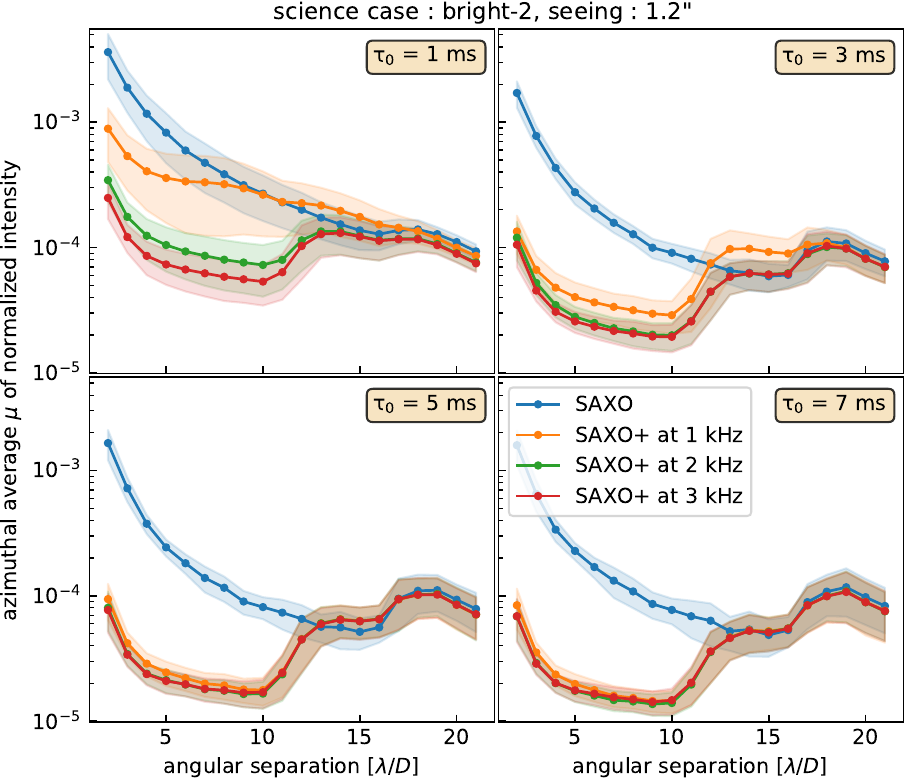}
    \caption{Bright-2, seeing = 1.2".}
    \label{fig:freq2_bright2_seeing_1.2}
\end{figure}

\begin{figure}
    \centering
    \includegraphics[width=0.95\hsize]{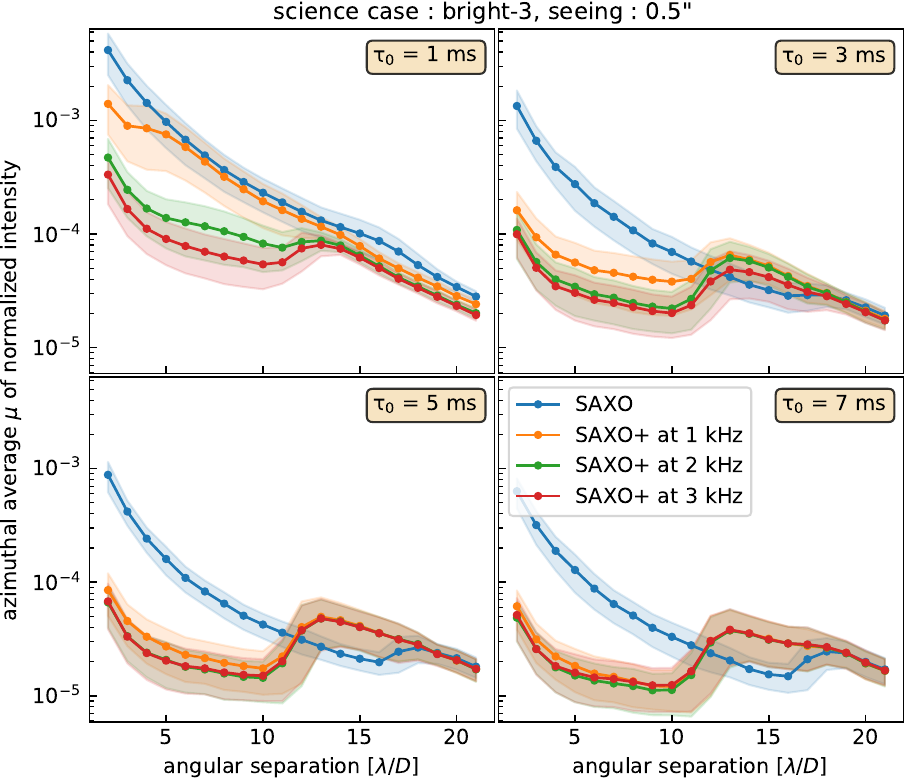}
    \caption{Bright-3, seeing = 0.5".}
    \label{fig:freq2_bright3_seeing_0.5}
\end{figure}

\begin{figure}
    \centering
    \includegraphics[width=0.95\hsize]{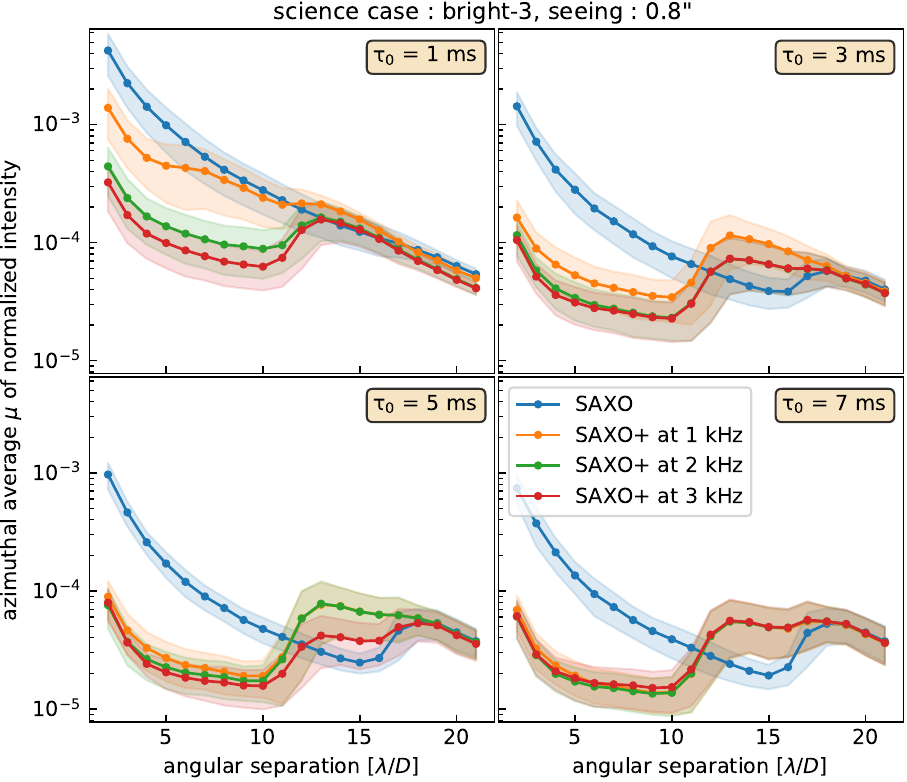}
    \caption{Bright-3, seeing = 0.8".}
    \label{fig:freq2_bright3_seeing_0.8}
\end{figure}

\begin{figure}
    \centering
    \includegraphics[width=0.95\hsize]{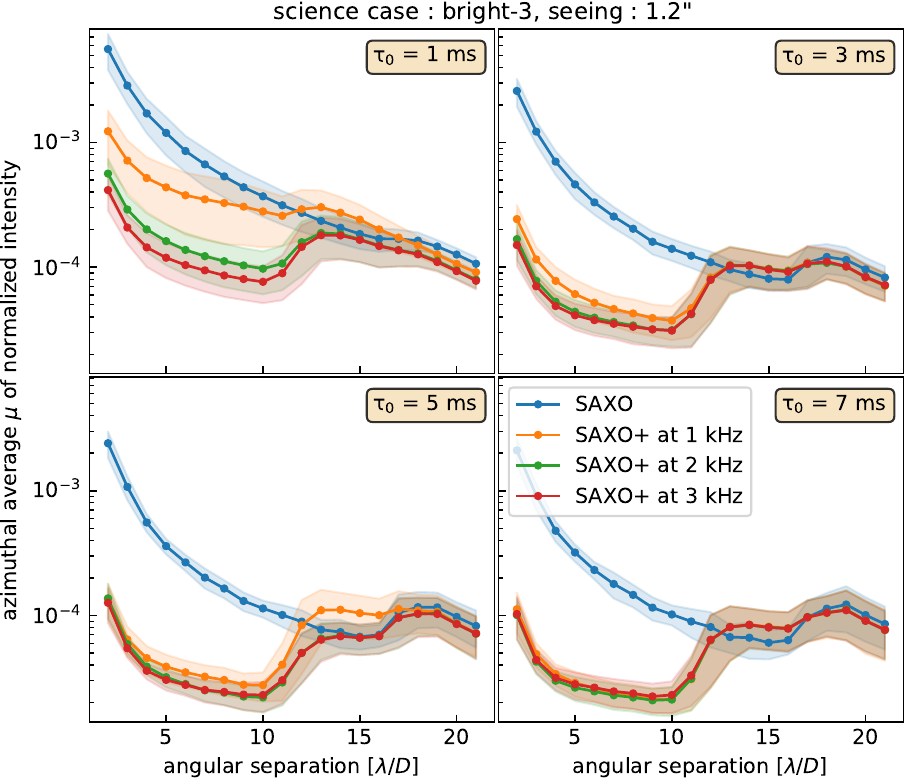}
    \caption{Bright-3, seeing = 1.2".}
    \label{fig:freq2_bright3_seeing_1.2}
\end{figure}

\begin{figure}
    \centering
    \includegraphics[width=0.95\hsize]{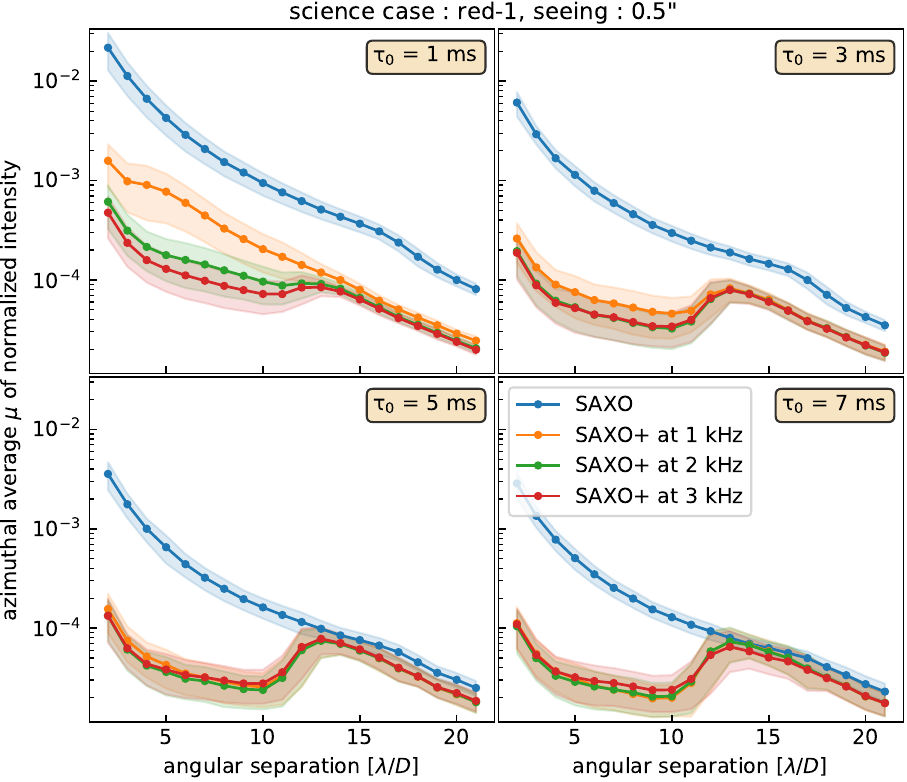}
    \caption{Red-1, seeing = 0.5".}
    \label{fig:freq2_red1_seeing_0.5}
\end{figure}

\begin{figure}
    \centering
    \includegraphics[width=0.95\hsize]{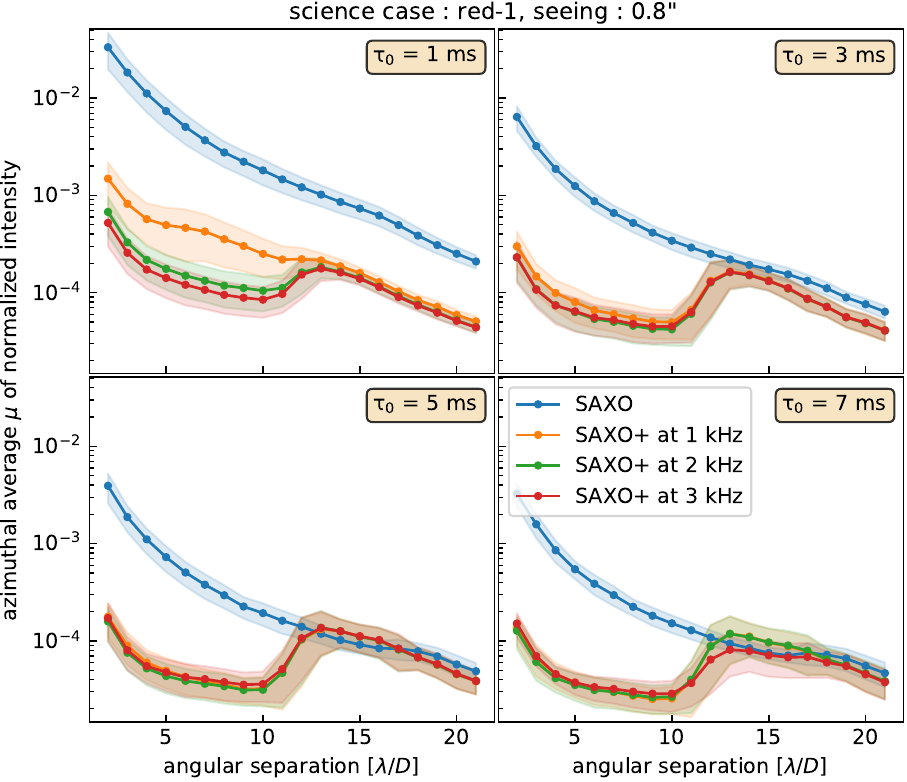}
    \caption{Red-1, seeing = 0.8".}
    \label{fig:freq2_red1_seeing_0.8}
\end{figure}

\begin{figure}
    \centering
    \includegraphics[width=0.95\hsize]{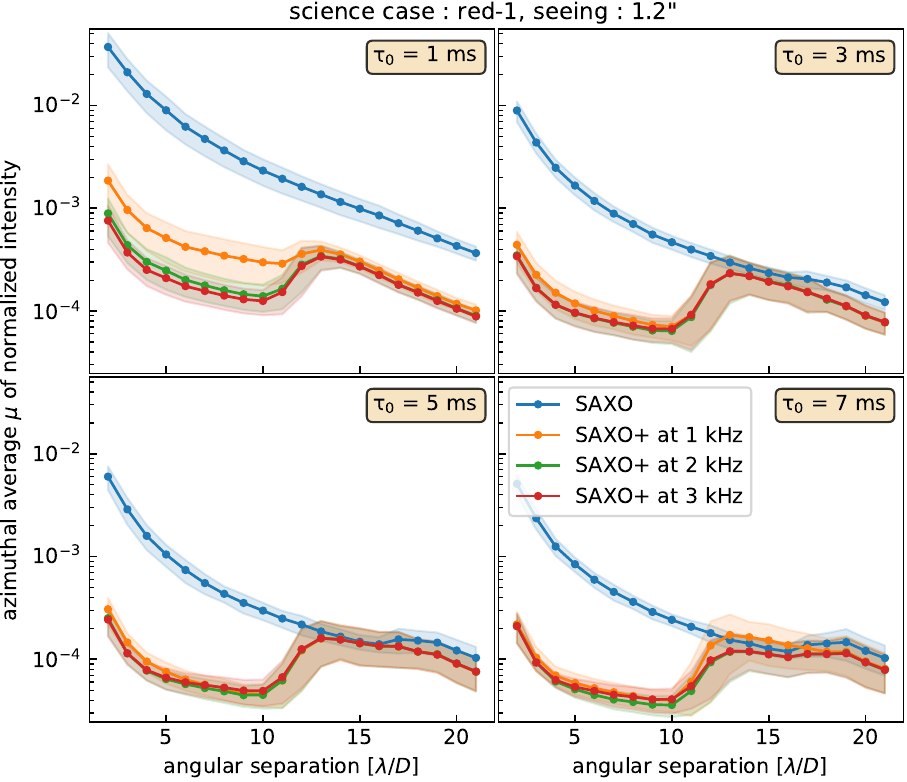}
    \caption{Red-1, seeing = 1.2".}
    \label{fig:freq2_red1_seeing_1.2}
\end{figure}

\begin{figure}
    \centering
    \includegraphics[width=0.95\hsize]{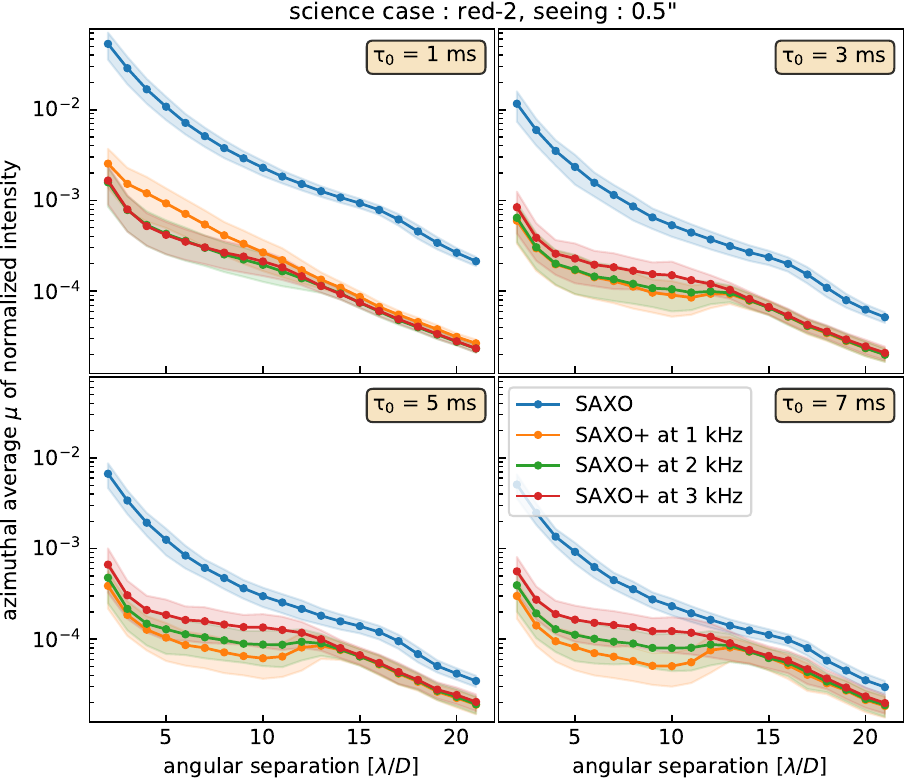}
    \caption{Red-2, seeing = 0.5".}
    \label{fig:freq2_red2_seeing_0.5}
\end{figure}

\begin{figure}
    \centering
    \includegraphics[width=0.95\hsize]{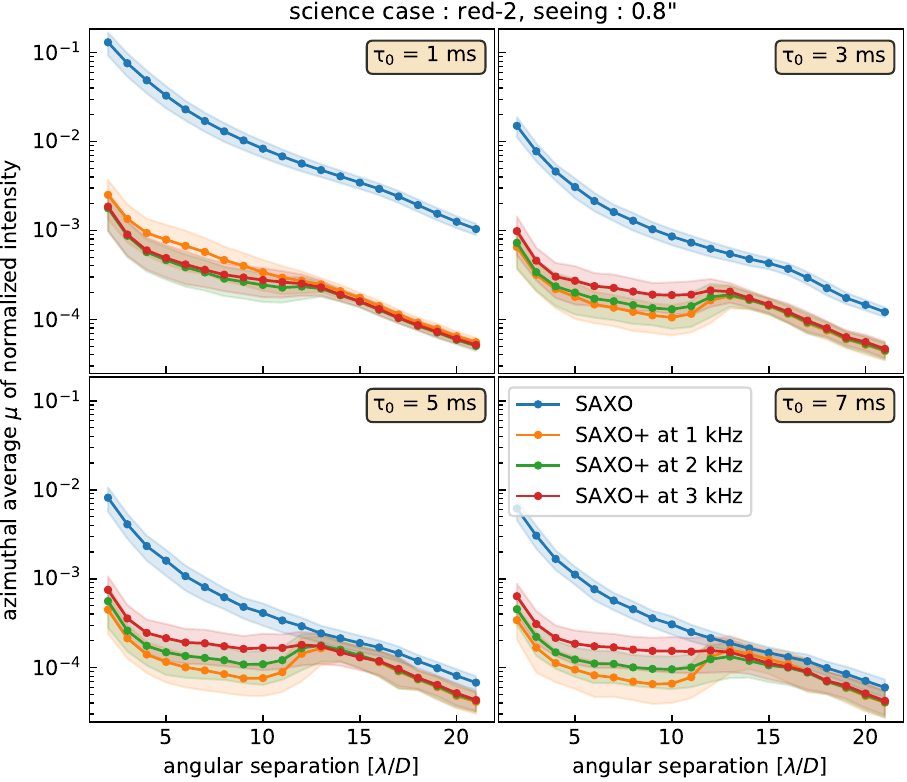}
    \caption{Red-2, seeing = 0.8".}
    \label{fig:freq2_red2_seeing_0.8}
\end{figure}

\begin{figure}
    \centering
    \includegraphics[width=0.95\hsize]{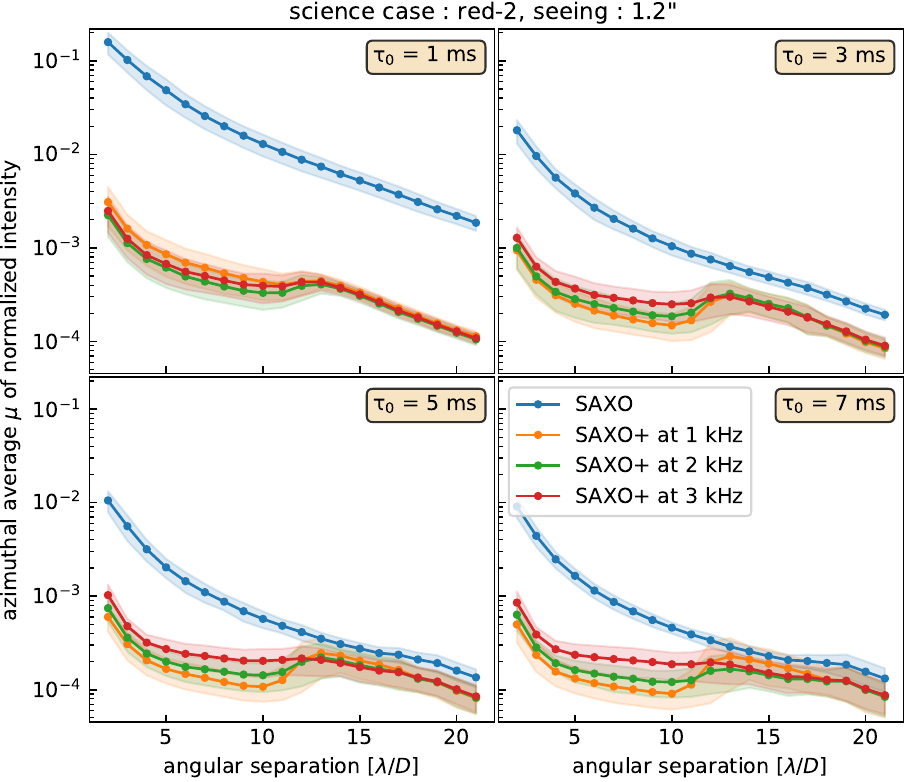}
    \caption{Red-2, seeing = 1.2".}
    \label{fig:freq2_red2_seeing_1.2}
\end{figure}

\begin{figure}
    \centering
    \includegraphics[width=0.95\hsize]{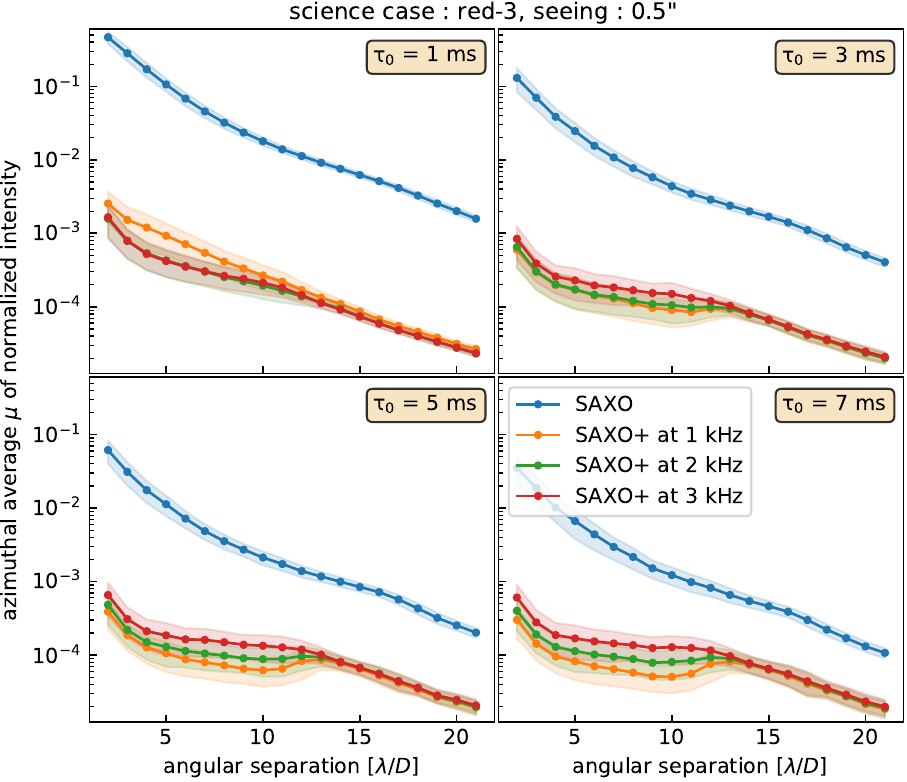}
    \caption{Red-3, seeing = 0.5".}
    \label{fig:freq2_red3_seeing_0.5}
\end{figure}

\begin{figure}
    \centering
    \includegraphics[width=0.95\hsize]{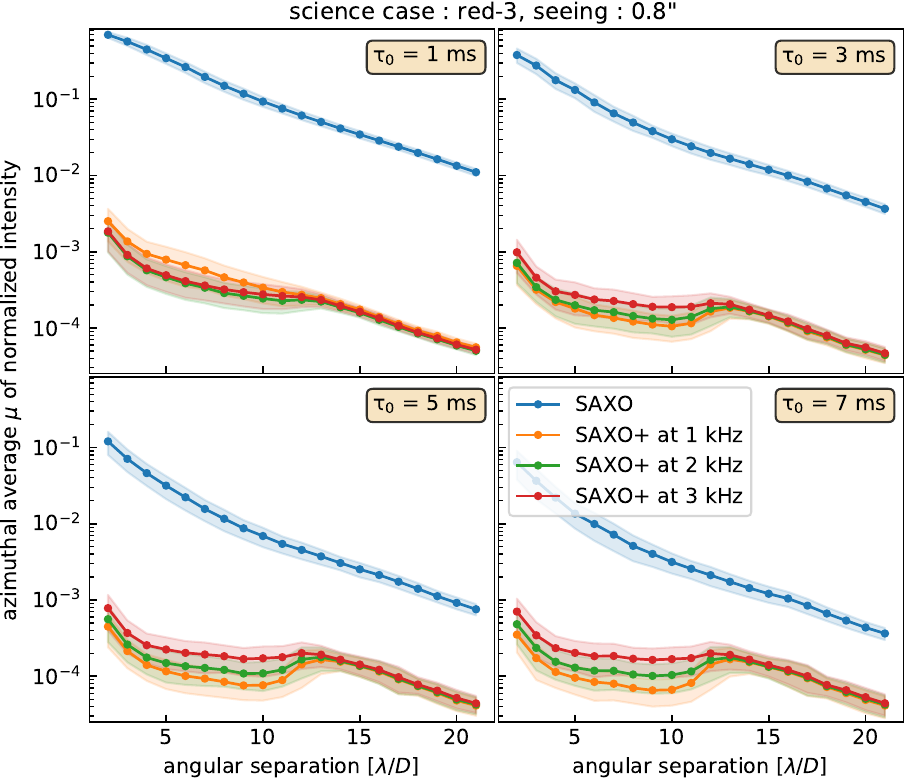}
    \caption{Red-3, seeing = 0.8".}
    \label{fig:freq2_red3_seeing_0.8}
\end{figure}

\begin{figure}
    \centering
    \includegraphics[width=0.95\hsize]{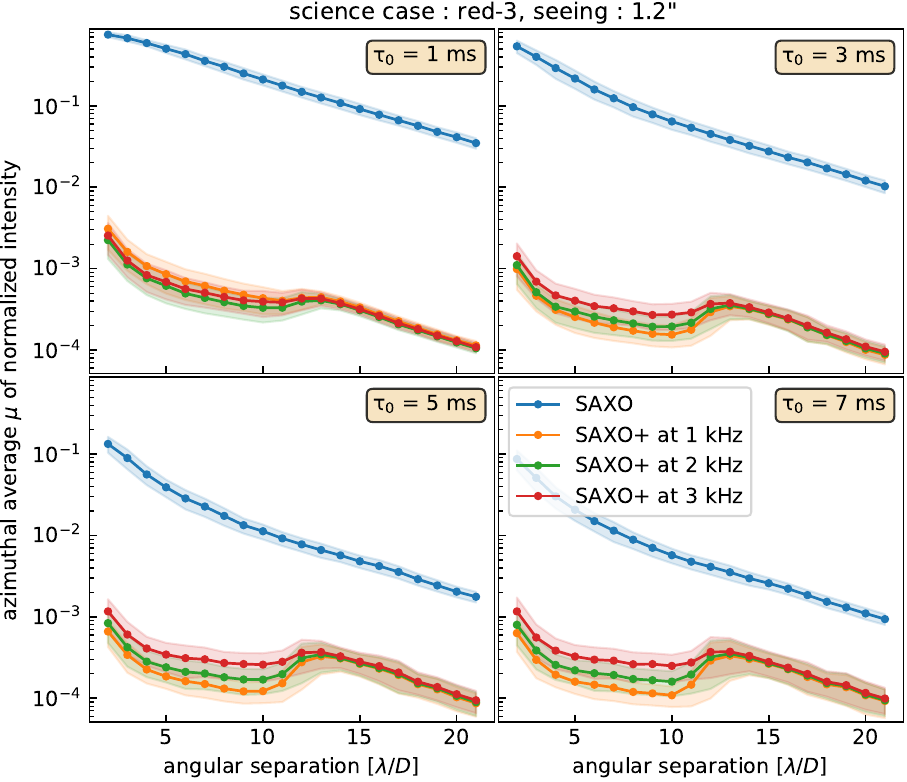}
    \caption{Red-3, seeing = 1.2".}
    \label{fig:freq2_red3_seeing_1.2}
\end{figure}

\begin{figure}
    \centering
    \includegraphics[width=0.95\hsize]{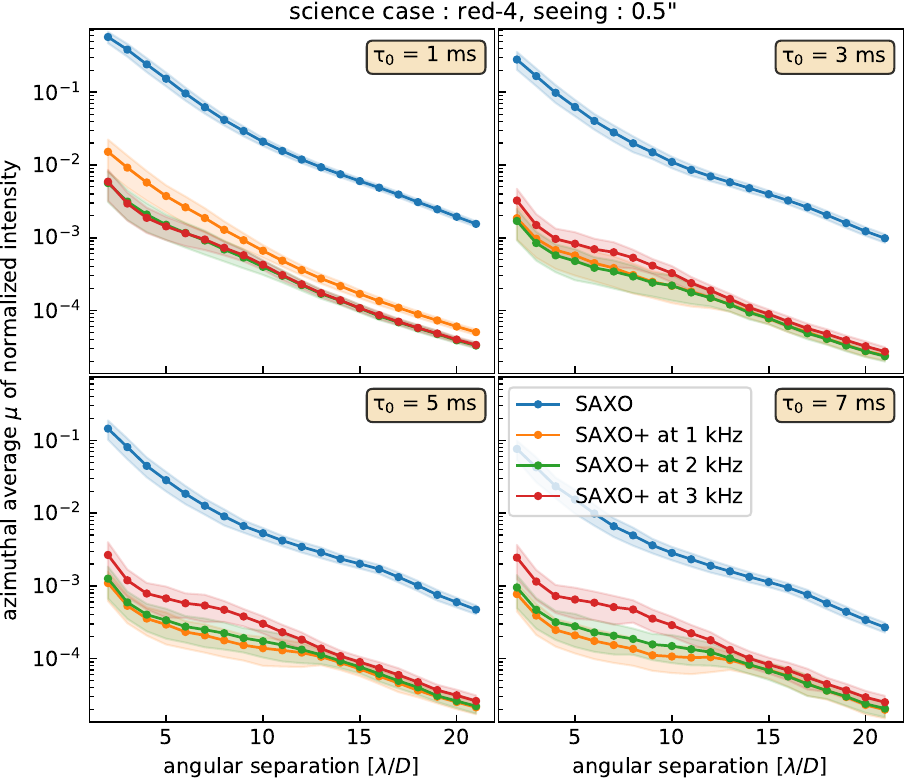}
    \caption{Red-4, seeing = 0.5".}
    \label{fig:freq2_red4_seeing_0.5}
\end{figure}

\begin{figure}
    \centering
    \includegraphics[width=0.95\hsize]{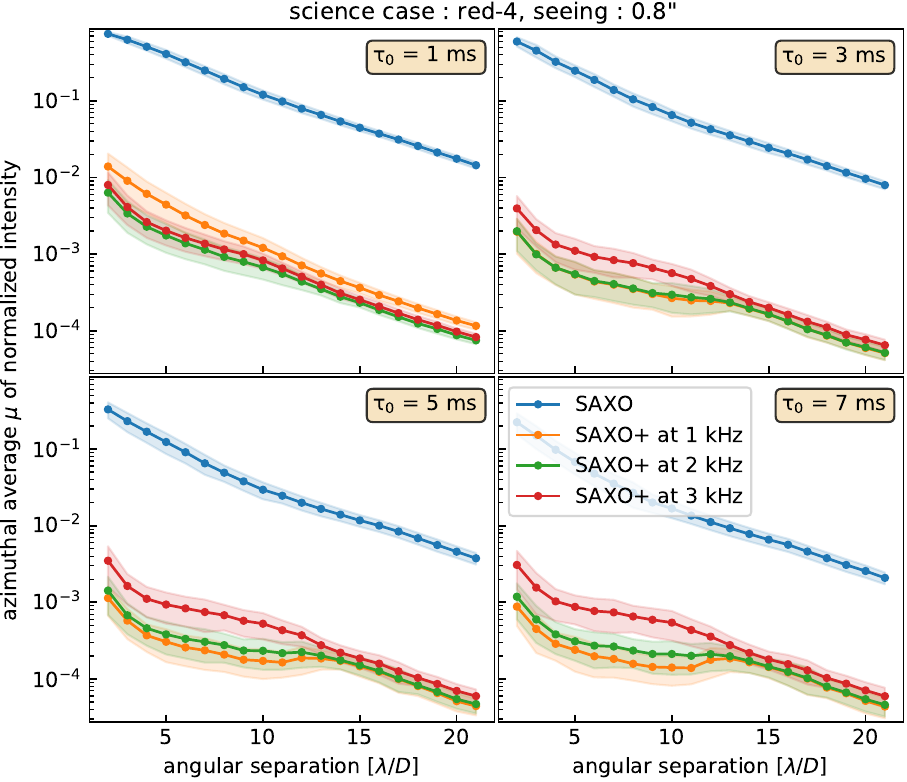}
    \caption{Red-4, seeing = 0.8".}
    \label{fig:freq2_red4_seeing_0.8}
\end{figure}

\begin{figure}
    \centering
    \includegraphics[width=0.95\hsize]{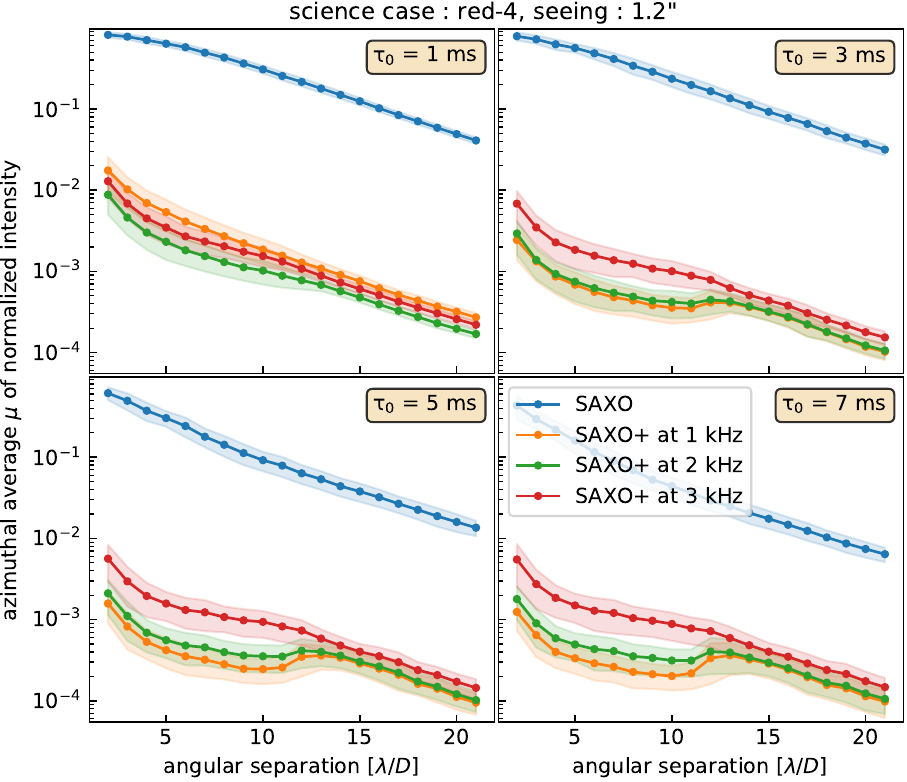}
    \caption{Red-4, seeing = 1.2".}
    \label{fig:freq2_red4_seeing_1.2}
\end{figure}

\begin{figure}
    \centering
    \includegraphics[width=0.95\hsize]{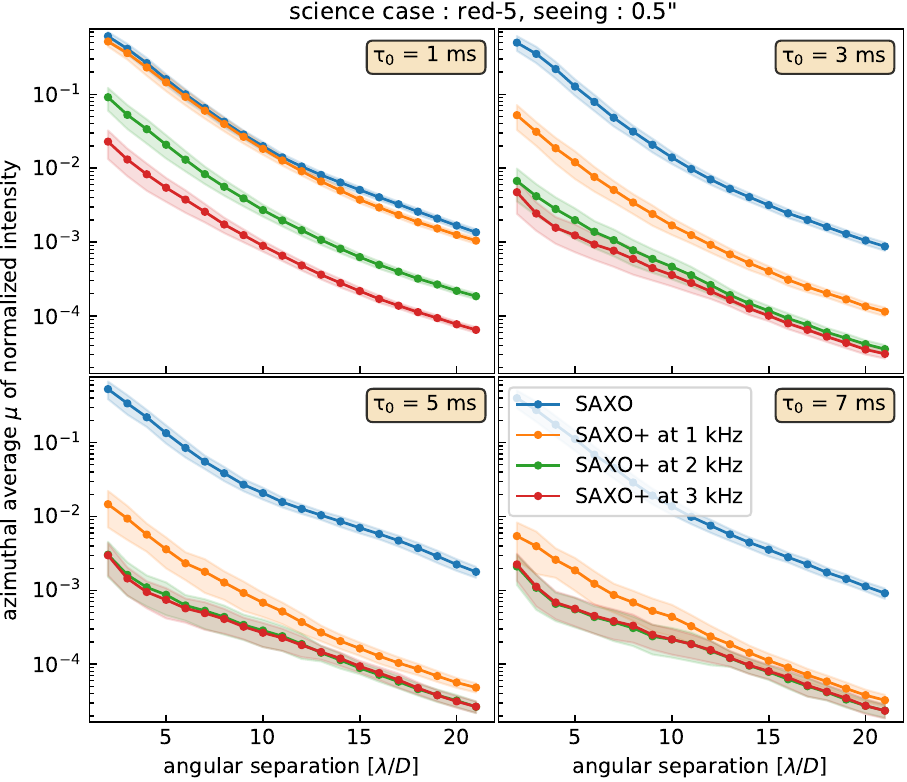}
    \caption{Red-5, seeing = 0.5".}
    \label{fig:freq2_red5_seeing_0.5}
\end{figure}

\begin{figure}
    \centering
    \includegraphics[width=0.95\hsize]{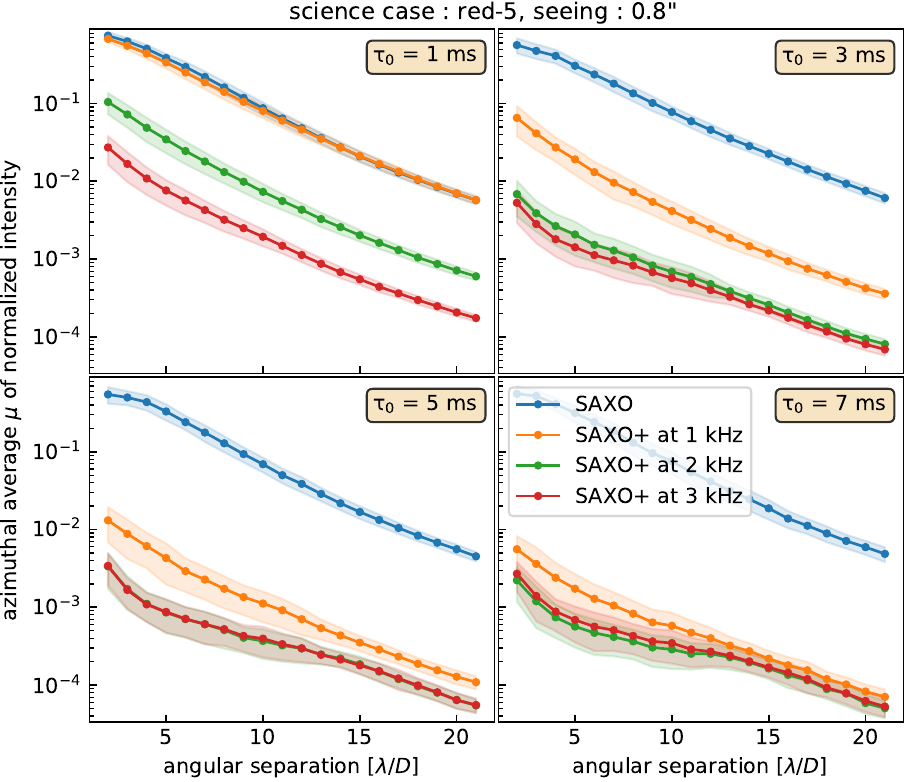}
    \caption{Red-5, seeing = 0.8".}
    \label{fig:freq2_red5_seeing_0.8}
\end{figure}

\begin{figure}
    \centering
    \includegraphics[width=0.95\hsize]{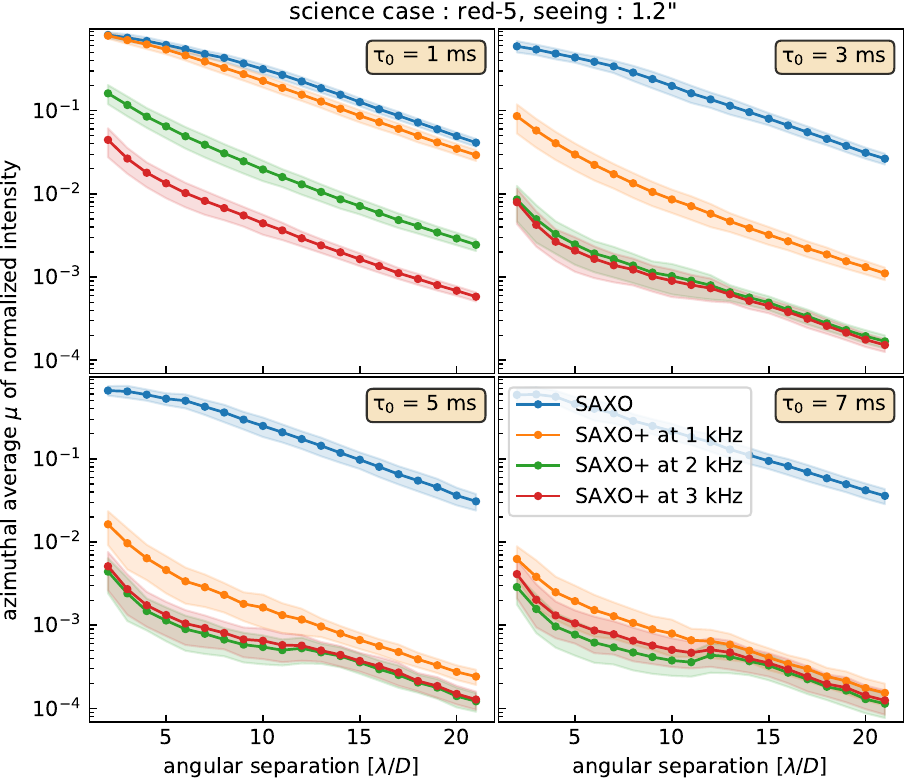}
    \caption{Red-5, seeing = 1.2".}
    \label{fig:freq2_red5_seeing_1.2}
\end{figure}

In Sect. \ref{subsec:2nd_freq} Fig. \ref{fig:freq2}, we plot the azimuthal average $\mu$ of the normalized intensity for different frequencies of the second stage. The seeing is set to 0.8" and the science case is bright-1. From Fig. \ref{fig:freq2_bright1_seeing_0.5} to Fig. \ref{fig:freq2_red5_seeing_1.2} we plot the same figure for every other combination of seeing and science case. The PWFS modulation radius is fixed at 3 $\lambda_{\mathrm{WFS}} / D$.

\end{appendix}

\end{document}